\documentclass[aps,prd,twocolumn,preprintnumbers,superscriptaddress, showpacs]{revtex4}

\usepackage{amsmath,amssymb}
\usepackage[dvips]{graphicx}
\usepackage{bm}

\DeclareMathOperator{\tr}{tr}

\DeclareMathOperator{\ring}{ring}

\newcommand{\Slash}[1]{{\ooalign{\hfil/\hfil\crcr$#1$}}}

\begin{document}
\preprint{\normalsize UT-Komaba/09-1} 


\title{
The strongly coupled fourth family and a first-order electroweak phase transition   
\\ (I) quark sector 
%
%
}


\author{Yoshio Kikukawa}
\email[]{kikukawa@hep1.c.u-tokyo.ac.jp}
\affiliation{Institute of  Physics, University of Tokyo\ 
Tokyo 153-8092, Japan}

\author{Masaya Kohda}
\email[]{mkohda@eken.phys.nagoya-u.ac.jp}
\affiliation{Department of Physics, Nagoya University\ 
Nagoya 464-8602, Japan}

\author{Junichiro Yasuda}
\email[]{yasuda@cshe.nagoya-u.ac.jp}
\affiliation{Center for the Studies of Higher Education, Nagoya University\ 
Nagoya 464-8601, Japan}


\date{\today}

\begin{abstract}
In models of dynamical electroweak symmetry breaking due to strongly coupled fourth-family 
quarks and leptons,  their low-energy effective descriptions may involve multiple composite Higgs 
fields, leading to a possibility that 
the electroweak phase transition at finite temperature is first order due to the 
Coleman-Weinberg mechanism. 
We examine the behavior of the electroweak phase transition 
based on 
the effective renormalizable Yukawa theory 
which consists of the fourth-family quarks and two SU(2)-doublet Higgs fields corresponding 
to the bilinear operators of the fourth-family quarks
 with/without imposing the compositeness condition. 
%
The strength of the first-order phase transition is estimated by using the finite-temperature effective potential at one-loop with the ring-improvement.
In the Yukawa theory without the compositeness condition, 
it is found 
that there is a parameter region where 
the first-order phase transition is strong enough for the electroweak baryogenesis with the experimentally acceptable 
Higgs boson and fourth-family quark masses.
On the other hand, 
when the compositeness condition is imposed, 
the phase transition turns out to be weakly first order, or possibly second order, 
although the result is rather sensitive to the details of the compositeness condition. 
Combining with the result of the Yukawa theory without the 
compositeness condition,  
it is argued that with the fourth-family quark masses 
in the range of 330-480 GeV, corresponding to the compositeness scale in the range of 1.0-2.3 TeV, 
the four-fermion interaction among the fourth-family quarks
does not lead to the strongly first-order electroweak phase transition.

\end{abstract}

\pacs{11.10.Wx, 12.60.Fr, 98.80.Cq.}

\maketitle


\section{Introduction\label{sec:introduction}}

The standard model (SM) can in principle fulfill  all three Sakharov conditions \cite{Sakharov:1973} for generating a baryon asymmetry in the universe \cite{Kuzmin:1985mm,Cohen:1990py,Cohen:1990it}.
%
The model fails, however,  for two reasons, 
to explain the value of the asymmetry required for the primordial nucleosynthesis \cite{Steigman:2005uz}, or the value measured through the cosmic microwave background \cite{Spergel:2006hy}. 
%
The first reason is that the CP violation from the Kobayashi-Maskawa mechanism \cite{Kobayashi:1973fv}, which nicely explains CP violation in K- and B-systems,  
is highly suppressed \cite{Shaposhnikov:1987tw,Farrar:1993sp,Farrar:1993hn,Gavela:1994dt,Gavela:1993ts,Huet:1994jb}. The second reason is that the electroweak phase transition (EWPT) is not strongly first order. 
The experimental lower bound on the Higgs mass,  $m_h > 114$ GeV \cite{:2003ih}, implies that there is no EWPT in the SM \cite{Kajantie:1996mn,Rummukainen:1998as,Csikor:1998eu,Aoki:1999fi}. Consequently, sphaleron-induced (B+L)-violating interactions are not sufficiently suppressed in the broken phase and wash out the baryon asymmetry. 
Therefore, if the physics at the electroweak scale is to explain the baryon asymmetry in the universe,  the better understanding of  the structure of the Higgs sector and of the source of CP violation would 
be required.

The fourth family is still a viable phenomenological possibility beyond the SM. 
The constraint from the invisible Z width is insignificant for the fourth-family neutrino being heavier than $m_Z/2$.  
Though the electroweak precision data \cite{Alcaraz:2006mx} give stringent constraint on the fourth family\cite{Holdom:1990tc,Peskin:1990zt,Peskin:1991sw,Golden:1990ig}, 
it is known that the data do not exclude their 
existence \cite{Gates:1991uu,Kniehl:1992ez,Holdom:1996bn,He:2001tp}. 
It was shown that there still remain a parameter region being consistent with all current experimental bounds\cite{Kribs:2007nz}. 

The existence of the fourth-family quarks accommodate the extra mixings 
and CP violating phases within 
the Cabibbo-Kobayashi-Maskawa scheme. 
It has been argued that the observed anomaly in B-CP asymmetries may be explained 
by the effect of the fourth-family quarks \cite{Hou:2005hd,Hou:2006zza,Hou:2006jy,Soni:2008bc}. 
Recently, it was pointed out that CP violation from these new phases could be large enough to explain the baryon asymmetry in the universe on the basis of the dimensional analysis using the Jarlskog 
invariants extended to four families\cite{Hou:2008xd}. 

The question is then whether the EWPT can be strongly first order with the fourth family:  
the mere addition of the fourth family to the SM is of no help in this respect, 
as long as the standard Higgs sector with the single SU(2) doublet is considered. (See, for example,  a recent study by Fok and Kribs \cite{Fok:2008yg}.) 
Carena et al. \cite{Carena:2004ha} has first discussed
the possibility of a first-order EWPT due to new heavy fermions coupled strongly to Higgs bosons.   
They found that 
some heavy and strongly-interacting bosonic fields are required both to stabilize the effective potential against the large effect of the heavy fermions and to cause a first-order EWPT 
\footnote{The authors would like to thank Y.~Okada and E.~Senaha for discussions on the result of \cite{Carena:2004ha}. }. 
This led the authors to consider a supersymmetric model. 
The EWPT in the supersymmetric model with the fourth family has recently been examined in 
\cite{Fok:2008yg}. (See \cite{Ham:2004xh} for earlier works.)


If the masses of the fourth-family quarks and leptons are quite large and are comparable to 
the unitarity bounds, the fourth family must couple strongly to the Higgs sector \cite{Chanowitz:1978mv}. 
In this case,  the masses of the fourth-family quarks and leptons (or their vacuum condensates)
may be regarded as the order parameters of the electroweak symmetry breaking (EWSB)
\footnote{The model of dynamical electroweak symmetry breaking due to top quark condensate 
has been proposed by Nambu and Miransky-Tanabashi-Yamawaki \cite{Nambu:1988mr,Miransky:1988xi,Miransky:1989ds} 
and examined in detail through renormalization group methods by Marciano and Bardeen-Hill-Lindner 
\cite{Marciano:1989mj, Marciano:1989xd,Bardeen:1989ds}.}
\cite{Nambu:1961tp,Holdom:1986rn,Holdom:1983kw, Holdom:1990ta,Hill:1990ge,Elliott:1992xg,Hill:1992ev}. 
The effective description of the fluctuations of the order parameters may 
involve multiple Higgs scalar fields. This leads to the possibility that the
EWPT would be first order due to the Coleman-Weinberg mechanism (the fluctuation induced first-order phase transition) 
\cite{Coleman:1973jx,Bak-Krinsky-Mukamel:1976,Rudnick:1978,Amit:1978dk,Ginsparg:1980ef,Iacobson:1981jm,Yamagishi:1981qq,Pisarski:1983ms,Chivukula:1992pm}
\footnote{The original Higgs sector of the SM, if the electroweak interactions is turned off,  
is nothing but the O(4) linear sigma model and
its finite temperature phase transition is second order,  which is governed by the 
Wilson-Fisher IR-stable fixed point.
It is the effect of the gauge interaction which makes the fixed point IR-unstable and causes a 
first-order phase transition \cite{Coleman:1973jx}. 
Once the Higgs sector is extended, the number of the scalar fields is increased and there appear 
additional quartic couplings among them. The fixed points of the multiple quartic coupling constants 
may be IR-unstable and 
one can expect a first-order phase transition even in the pure scalar 
sector \cite{Bak-Krinsky-Mukamel:1976,Iacobson:1981jm}. 
A related approach is to consider the single Higgs doublet model (the O(4) model) with the dimension-six or higher operators \cite{Grojean:2004xa,Ham:2004zs,Bodeker:2004ws,
Delaunay:2007wb,Grinstein:2008qi}. 
The quartic coupling may  be then assumed negative so that the model is out of the domain of the Wilson-Fisher  fixed point.  
Such higher dimensional operators may be induced by the effect of heavy particles coupled to the 
Higgs doublet, or more generally, by the effect of a certain dynamical system behind the Higgs sector. 
The EWPT has been examined in various dynamical models of the Higgs sector:
walking technicolor theories in \cite{Kikukawa:2007zk, Cline:2008hr,Jarvinen:2009wr} 
using low-energy effective sigma models, 
pseudo Goldstone Higgs boson models or little Higgs models in \cite{Grinstein:2008qi,Espinosa:2004pn}, models of the gauge-Higgs unification in \cite{Panico:2005ft,Maru:2005jy,Maru:2006wx}. 
%
}.

The goal of this paper is to explore the above possibility of a first-order EWPT due to the 
heavy fourth family.  
We start from a model of dynamical electroweak symmetry breaking 
due to the effective four-fermion interactions
of the fourth-family quarks and leptons  
at the scale $\Lambda_{\rm 4f}$ around a few TeV.  
We adopt the four-fermion interactions considered by Holdom \cite{Holdom:2006mr}. 
This four-fermion theory may be rewritten 
into a Yukawa theory 
by introducing auxiliary scalar fields  which corresponds to the bilinear operators of 
the fourth-family quarks and leptons. 
These scalar fields consist of three SU(2) doublets and one SU(2) triplet.
(It is assumed that the right-handed neutrino is extra heavy,  acquiring 
its mass at the flavor scale around 1000 TeV. )
The renormalization group evolution from the scale $\Lambda_{\rm 4f}$ down to the electroweak scale 
$v$ (= 246 GeV) may generate operators such as the 
kinetic and interaction terms of the scalar fields and other higher dimensional operators. 
We then extend this model 
by  including the kinetic, cubic and quartic terms of the scalar fields 
so that it becomes {\em renormalizable}, neglecting the effect of the higher dimensional operators. 
It is this effective renormalizable model for which we examine EWPT
through the finite-temperature effective potential at one-loop with the ring-improvement 
\cite{Dolan:1973qd, Weinberg:1974hy, Anderson:1991zb, 
Fendley:1987ef, Espinosa:1992gq, Parwani:1991gq, Carrington:1991hz, Dine:1992wr, Arnold:1992rz, Fodor:1994bs}. 
Strictly speaking, in our case,
the renormalization group equations must be subject to the compositeness condition 
as a boundary condition 
at the scale $\mu=\Lambda_{\rm 4f}$ \cite{Bardeen:1989ds,Harada:1994wy}. Accordingly,  
the values of the renormalized couplings at the lower scale $\mu=v$ are restricted in a certain region
of the parameter space.  In our analysis, however, we will first explore the parameter space of the renormalizable theory without the constraint due to the compositeness condition, 
in order to locate  the parameter region where a strongly first-order EWPT is realized. 
We then examine the possible overlap of these two regions. 

In this paper (I), we concentrate on the effect of the heavy quarks and consider two 
SU(2) doublets out of four scalar fields.  
The bosonic sector of our model then reduces to the two Higgs doublet 
model (2HDM) \cite{Bochkarev:1990fx,Bochkarev:1990gb,Cohen:1991iu,
Nelson:1991ab,Turok:1990zg,Turok:1991uc,Funakubo:1993jg,Davies:1994id,Cline:1995dg,
Cline:1996mg,Fromme:2006cm,Kanemura:2004ch,Aoki:2008av}.
The analysis of the effect of the heavy charged lepton 
and neutrino will be reported in a subsequent paper \cite{Kikukawa-Kohda:2009xx}. 
We also neglect,  in this paper, the SU(3)$\times$SU(2)$\times$U(1) gauge interaction and consider 
the global symmetry limit 
because we do not expect a large effect 
of the electroweak interaction to the dynamics of the first-order phase transition 
in this model \cite{Kikukawa:2007zk}.  

This paper is organized as follows. In section~\ref{sec:fourth-family-EWSB}, 
we formulate  the effective Yukawa theory and 
introduce the cutoff scale $\Lambda$ by considering the vacuum instability and the triviality bounds.  
We then specify  the compositeness condition for our model. 
In section~\ref{sec:effective-potential-at-one-loop},  we derive the finite-temperature 
effective potential at one-loop with the ring-improvement. 
In section~\ref{sec:numerical-results},  based on the numerical analysis of the effective potential, 
we examine the strength of the first-order phase transition in the Yukawa theory with/without the 
compositeness condition.
Section~\ref{sec:conclusion-discussion} is devoted to conclusion and discussion. 

\section{Fourth family and Electroweak symmetry breaking}
\label{sec:fourth-family-EWSB}

\subsection{Fourth family and Four-fermion interactions}

We assume the existence of  the fourth-family quarks and leptons, which we denote by 
$q^\prime =(t^\prime , b^\prime)^T$, 
$\ell^\prime_L =(\nu^\prime_{\tau L} , \tau^\prime_L)^T$, $\tau^\prime_R$. 
The right-handed neutrino $\nu^\prime_{\tau R}$ is assumed to acquire its mass at the flavor scale around 1000 TeV and to be absent  below the flavor scale.
To be consistent with the electroweak precision date, the masses of the fourth-family quarks should be almost degenerate with a small mass splitting.
For simplicity, we assume $m_{t^\prime}=m_{b^\prime}$. 

Following Holdom\cite{Holdom:2006mr}, we introduce the four-fermion interactions of the fourth-family fermions as follows:
\begin{align}
\mathcal{L}_{\rm 4f} =&
G_{q^\prime}(\bar{q^\prime}_{Li} q_{Rj}^\prime)(\bar{q^\prime}_{Rj} q_{Li}^\prime)
+G_{\tau^\prime}(\bar{\ell^\prime}_{Li} \tau_{R}^\prime)(\bar{\tau^\prime}_{R} \ell_{Li}^\prime) \notag\\
&-G_{\nu_{\tau L}^\prime}(\ell_{Li}^{\prime T} C^{\dagger}\ell_{Lj}^\prime)(\bar{\ell^\prime}_{Lj} C  \bar{\ell^\prime}_{Li}^T), 
 \label{eq:4fermion}
\end{align}    
where $C$ is the charge conjugation matrix and color indexes are contracted within a bracket. 
The scale of these interactions is assumed to be $\Lambda_{\rm 4f}$: $G_{q^\prime}, G_{\tau^\prime}, G_{\nu_{\tau L}^\prime} \simeq 1/ \Lambda_{\rm 4f}^2$. 
The interaction term among the quarks has 
SU(2)$_L\times$SU(2)$_R\times$U(1)$_V\times$U(1)$_A$ symmetry, 
where U(1)$_V$ 
corresponds to the baryon number. 
On the other hand, the interaction terms among the leptons have 
SU(2)$_L\times$U(1)$_{L}\times$U(1)$_R$ 
symmetry which includes 
the  vector like U(1) symmetry corresponding to the lepton number.
The above four-fermion interactions, therefore,  have the extra symmetries compared with the SM Higgs sector which has O(4) symmetry. 
We then 
assume the existence of sub-leading multi-fermion operators, which are suppressed compared with 
Eq.~(\ref{eq:4fermion}), so that 
the extra symmetries are broken explicitly and, hence, the possible pseudo Nambu-Goldstone (NG) bosons acquire 
non-zero masses. 

The four-fermion interactions may be rewritten into the form of Yukawa interactions by introducing the 
auxiliary scalar fields
$\Phi$, $H_{\tau^\prime}$ and $\chi^a$ ($a=1,2,3$), which correspond  to the bilinear operators of 
the fourth-family quarks and leptons as follows:
\begin{align}
\Phi_{ij}\sim \bar{q^\prime}_{Rj} q_{Li}^\prime ~,~ H_{\tau^\prime}\sim \bar{\tau^\prime}_{R} \ell_{Li}^\prime ~,~
\chi^a \sim \bar{\ell^\prime}_{L} \tau^a \epsilon C \bar{\ell^\prime }_{L}^T . 
\end{align} 
Then one obtains
\begin{align}
\label{eq:four-fermion-theory-equivalent}
\mathcal{L}_{\rm 4f}^\prime=-m_{\Phi 0}^2\tr(\Phi^\dagger\Phi) - m_{H_{\tau^\prime} 0}^2 H_{\tau^\prime}^\dagger H_{\tau^\prime}
-m_{\chi 0}^2 \chi^{a*}\chi^a
+\mathcal{L}_{\rm Y},
\end{align}
where 
$\mathcal{L}_{\rm Y}$ is the Yukawa interaction terms given by
\begin{align}
\mathcal{L}_{\rm Y} =& -y_{q^\prime 0}(\bar{q^\prime}_{L} \Phi q_{R}^\prime + c.c.) 
-y_{\tau^\prime}(\bar{\ell^\prime}_{L} H_{\tau^\prime} \tau_{R}^\prime +c.c.) \notag \\
&-f(\ell_{L}^{\prime T} C^{\dagger}\epsilon \tau^a \chi^a \ell_{L}^\prime +c.c.). 
\end{align}

\subsection{Effective Renormalizable Theory}

Through the renormalization group evolution from the scale $\mu=\Lambda_{\rm 4f}$ 
down to the electroweak scale $\mu=v$ (=246 GeV),  the kinetic and 
interaction terms of the scalar fields and other higher dimensional operators may be generated. 
We then extend this model 
by  including the kinetic, cubic and quartic terms of the scalar fields 
so that it becomes  renormalizable, {\em neglecting the effect of the higher dimensional operators}. 
The effective renormalizable theory is then given by the following Lagrangian:
\begin{align}
\label{eq:effctive-renormalizable-model-for-fourth-family}
\mathcal{L}=\mathcal{L}_{\rm k}+\mathcal{L}_{\rm Y}-V,
\end{align}
where $\mathcal{L}_{\rm k}$ consists of the kinetic terms for fourth-family fermions and scalar bosons
and $V$ is the scalar potential.
The explicit form of $V$ is given in appendix \ref{sec:appendix} 
\footnote{$V$ includes a scalar cubic term $H_{\tau^\prime}^T\epsilon \tau^a\chi^{a*}H_{\tau^\prime}$ which may enhance a strength of a first-order phase transition at high temperature.}.

Strictly speaking, 
the renormalization group equations are subject to the compositeness condition 
as a boundary condition at the scale $\mu=\Lambda_{\rm 4f}$ \cite{Bardeen:1989ds}. 
Accordingly, the values of the renormalized couplings at the lower scale $\mu=v$ are restricted in a certain region
of the parameter space.  In the following analysis, however, we will first explore the parameter space of 
the renormalizable theory without the constraint due to the compositeness condition, 
in order to locate  the parameter region where a strongly first-order EWPT is realized. 
We then examine the possible overlap of these two regions. 

In this paper (I), we concentrate on the effect of the fourth-family quarks and consider two 
SU(2) doublets out of four scalar fields.
We also neglect,  in this paper, the SU(3)$\times$SU(2)$\times$U(1) gauge interaction and consider 
the global symmetry limit, 
simply because we do not expect a large effect 
of the color and the electroweak interactions on the dynamics of the first-order phase transition 
in this model.  
Then the Lagrangian Eq.~(\ref{eq:effctive-renormalizable-model-for-fourth-family}) reduces to 
\begin{eqnarray}
\label{eq:lagrangian}
\mathcal{L} &= &
\bar{q^\prime}i\Slash{\partial} q^\prime -y(\bar{q^\prime}_L\Phi  q^\prime_R+c.c.) \nonumber\\
&+&
\tr (\partial _\mu \Phi ^\dagger \partial ^\mu \Phi ) - m_{\Phi}^{2} \tr\Phi ^\dagger \Phi \notag\\
&-& 
  \frac{\lambda _1}{2} (\tr\Phi ^\dagger \Phi)^2 - \frac{\lambda _2}{2}\tr (\Phi ^\dagger \Phi )^2 
+ c(\det\Phi+c.c.). 
 \label{eq:Llsm}
\end{eqnarray}
We include the last term which breaks the U(1)$_A$ symmetry and induces the mass of the pseudo NG boson. 
Then, the symmetry of the theory is the chiral symmetry SU(2)$_L\times$SU(2)$_R$ plus the U(1)$_V$ 
symmetry corresponding 
to the baryon number.
We do not include the terms which consist of $\epsilon \Phi^* \epsilon$ other than in the determinant term.

\subsection{Electroweak Symmetry Breaking}

We assume that the chiral symmetry SU(2)$_L\times$SU(2)$_R$ breaks down to 
the diagonal subgroup SU(2)$_V$ by 
the vacuum expectation value (VEV) of $\Phi(x)$:
\begin{equation}
\langle \Phi \rangle = \frac{\phi}{\sqrt{2N_f}} \mathbf{I} , 
\end{equation}
where $N_f(=2)$ is the number of the fourth-family quark flavors, $\mathbf{I}$ is the $N_f \times N_f$ unit matrix and $\phi \geq 0$. 
At tree-level, the VEV is determined by the effective potential: 
\begin{equation}
\label{eq:eff-potential-tree}
V_0(\phi)=\frac{1}{2}(m_\Phi^2-c)\phi^2+\frac{1}{8}\left( \lambda_1+\frac{\lambda_2}{N_f} \right)\phi^4 .
\end{equation}
For $(m_{\Phi }^2 -c)< 0$, the VEV is given by 
\begin{equation}
\phi_0=\sqrt{\frac{-2(m_\Phi^2-c)}{\lambda_1+\lambda_2/N_f}}. 
\end{equation}
For the effective potential  to be stable in this channel, the following conditions must be satisfied:
\begin{equation}
{\lambda}_1+{\lambda}_2/N_f \ge 0, \quad {\lambda}_2 \ge 0 . 
\end{equation}

Around the VEV, we may parametrize the fluctuation of $\Phi(x)$ as follows:
\begin{equation}
\Phi(x) = \frac{ \phi + h + i \eta }{\sqrt{2N_f}}\mathbf{I}
+ \sum_{\alpha=1}^{3} ( \xi^\alpha + i \pi^\alpha ) \frac{\sigma^\alpha}{2} , 
\end{equation}
where $\sigma^\alpha$ $(\alpha=1,2,3)$ are the Pauli matrices. 
The fields $h, \eta, \xi^\alpha, \pi^\alpha$ and $q^\prime$ acquire masses at tree level as summarized in Table~\ref{tab:lsmm}, where, for notational simplicity, we use the following abbreviations:
\begin{eqnarray}
a_h&=&\frac{3}{2}(\lambda_1+\lambda_2/N_f),~
a_\xi =\frac{1}{2}(\lambda_1+3\lambda_2/N_f),  \\
a_\eta&=&a_\pi=\frac{1}{2}(\lambda_1+\lambda_2/N_f),~a_{q^\prime}=\frac{1}{2N_f}y^2, 
\end{eqnarray}
and
\begin{eqnarray}
b_h &=& a_h - a_\pi = (\lambda_1+\lambda_2/N_f), \\
b_\xi &=& a_\xi - a_\pi=(\lambda_2/N_f). 
\end{eqnarray}
The bosonic sector of this model is just the 2HDM. 
$h$ is the singlet of SU(2)$_V$ and corresponds to the SM Higgs boson.
The adjoint $\pi^\alpha$ are the NG bosons of the breaking of SU(2)$_L\times$SU(2)$_R$, while the singlet $\eta$ is 
the pseudoscalar Higgs boson and is also the pseudo NG boson associated with the breaking of the U(1)$_A$ symmetry. 
The adjoint $\xi^\alpha$ consist of the extra neutral Higgs boson and the charged Higgs bosons.
Three NG bosons $\pi^\alpha$ are eaten by W and Z bosons when the electroweak interactions are introduced.   
As for the fourth-family quarks, the experimental lower bound from the direct search $m_{q^\prime}\gtrsim 256$ GeV 
\cite{Amsler:2008zz} implies $y\gtrsim 2.1$ at tree-level by taking $\phi_0=v$(= 246GeV).

\begin{table}
\begin{center}
  \begin{tabular}{cccc} \hline
    particle & $m_i^2(\phi)$ & $m_i^2(\phi_0)$ & $n_i$ \\ \hline
    h & $m_\Phi^2-c+a_h \phi^2$ & $b_h \phi_0^2$ & 1 \\
    $\xi$ & $m_\Phi^2+c+a_\xi \phi^2$ & $2c+b_\xi \phi_0^2$ & 3 \\
    $\eta$ & $m_\Phi^2+c+a_\eta \phi^2$ & 2c & 1 \\ 
    $\pi$ & $m_\Phi^2-c+a_\pi \phi^2$ & 0 & 3 \\ 
    $q^\prime$ & $ a_{q^\prime}\phi^2$ & $a_{q^\prime}\phi_0^2$ & $-24$ \\ \hline
\end{tabular}
\caption{The effective masses and the numbers of the degrees of freedom in the model for the fourth family quarks}
\label{tab:lsmm}
\end{center}
\end{table}

\subsection{Cutoff Scale of the Effective Theory}

\begin{figure}[b]
\begin{center}
\includegraphics[width=7cm,clip]{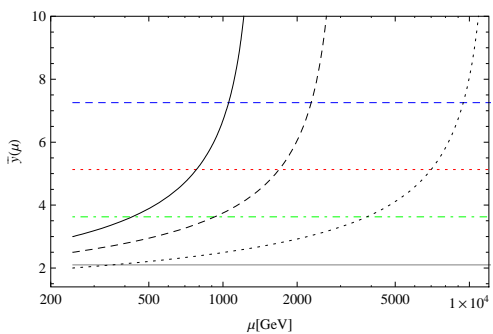}
\caption{The renormalization group flows of the Yukawa coupling for various initial values at the electroweak scale. The values of $y$ are 2.0, 2.5, 3.0 from bottom to top. 
The gray solid line indicates the value $y=2.1$, 
which corresponds to the experimental lower bound 
from the direct search of the fourth-family quarks, 
$m_{q^\prime}\gtrsim 256$ GeV. 
The blue dashed line indicates the upper limit of the perturbativity condition, which is adopted in our analysis as a criterion for the Landau pole.
}
\label{yukawa-running}
\end{center}
\end{figure}

The applicability of  the effective theory 
defined by the Lagrangian Eq.~(\ref{eq:lagrangian}) would break down at some energy scale 
and one needs to introduce a cutoff $\Lambda$. 
The running coupling constants in this model, $\bar{\lambda}_1(\mu)$, $\bar{\lambda}_2(\mu)$ and $\bar{y}(\mu)$,  obey
the following renormalization group equations at one-loop:
\begin{align}
\mu\frac{\partial}{\partial\mu}\bar{\lambda}_1&=\frac{1}{8\pi^2}[(N_f^2+4)\bar{\lambda}_1^2
+4N_f\bar{\lambda}_1\bar{\lambda}_2+3\bar{\lambda}_2^2+2N_c y^2\bar{\lambda}_1] ,  \notag \\
\mu\frac{\partial}{\partial\mu}\bar{\lambda}_2&=\frac{1}{8\pi^2}(6\bar{\lambda}_1\bar{\lambda}_2+2N_f\bar{\lambda}_2^2+2N_c y^2\bar{\lambda}_2-2N_c \bar{y}^4 ) ,  \notag \\
\mu\frac{\partial}{\partial\mu}\bar{y}&=\frac{1}{16\pi^2}(N_f+N_c)\bar{y}^3,
\label{eq:RG-eq-yukawa}
\end{align}
with the initial conditions $\bar{\lambda}_1(v)=\lambda_1$, $\bar{\lambda}_2(v)=\lambda_2$ and $\bar{y}(v)=y$ given at the electroweak scale.
As one can see in FIG.~\ref{yukawa-running}, 
the Yukawa coupling, which is large at the electroweak scale for the heavy fourth-family quarks,  
would blow up to infinity (due to the Landau pole) at a certain energy scale not far 
from the electroweak scale \footnote{If SU(3)$\times$SU(2)$\times$U(1) gauge interactions are 
included, there appears an effective infra-red fixed point in the renormalization group equation
of the Yukawa coupling Eq.~(\ref{eq:RG-eq-yukawa}) \cite{Hill:1980sq,Hill:1985tg}. 
But its value is about $y_c \simeq 1.6$,
which is less than the values of $y$ considered in our analysis.  So we neglect the effect of the 
infra-red fixed point. }. 
The ultraviolet behaviors of the scalar quartic couplings then take two types of pattern  
depending on the relative size of the scalar quartic couplings and Yukawa coupling at the electroweak scale:
(i) the scalar quartic coupling $\bar{\lambda}_1(\mu)+\bar{\lambda}_2(\mu)/N_f$ and/or $\bar{\lambda}_2(\mu)$ are driven to negative at some energy scale, implying that the electroweak vacuum is unstable; 
(ii) the scalar quartic couplings encounter the Landau pole at some energy scale. 
In both cases,  one should introduce a cutoff before these problems happen. 

We estimate the cutoff $\Lambda$, for given initial values of the couplings at the electroweak scale, 
as the scale at which one of the following conditions is first hit:
\begin{align}
\bar{\lambda}_1(\Lambda)+\bar{\lambda}_2(\Lambda)/N_f=0,  \quad \bar{\lambda}_2(\Lambda)=0,
\end{align}
which correspond to the case (i) (the vacuum instability) and   
\begin{eqnarray}
&& \bar{y}(\Lambda)^2= \frac{16\pi^2}{N_c}, \\
&& 
\bar{\lambda}_1(\Lambda)+\bar{\lambda}_2(\Lambda)/N_f= \frac{16\pi^2}{N_f^2}, \quad
\bar{\lambda}_2 (\Lambda)= \frac{16 \pi^2}{N_f},
\end{eqnarray}
which correspond to the case (ii) (the Landau pole).
Here, we adopt the upper limits of the perturbativity bounds, 
\begin{eqnarray}
&& \bar{y}(\Lambda)^2 \le \frac{16\pi^2}{N_c}, \\
&& 
\bar{\lambda}_1(\Lambda)+\bar{\lambda}_2(\Lambda)/N_f \le \frac{16\pi^2}{N_f^2}, \quad
\bar{\lambda}_2 (\Lambda)\le \frac{16 \pi^2}{N_f},
\end{eqnarray}
as a criterion for the Landau pole. 


\begin{figure}[t]
\begin{center}
\includegraphics[width=6cm,clip]{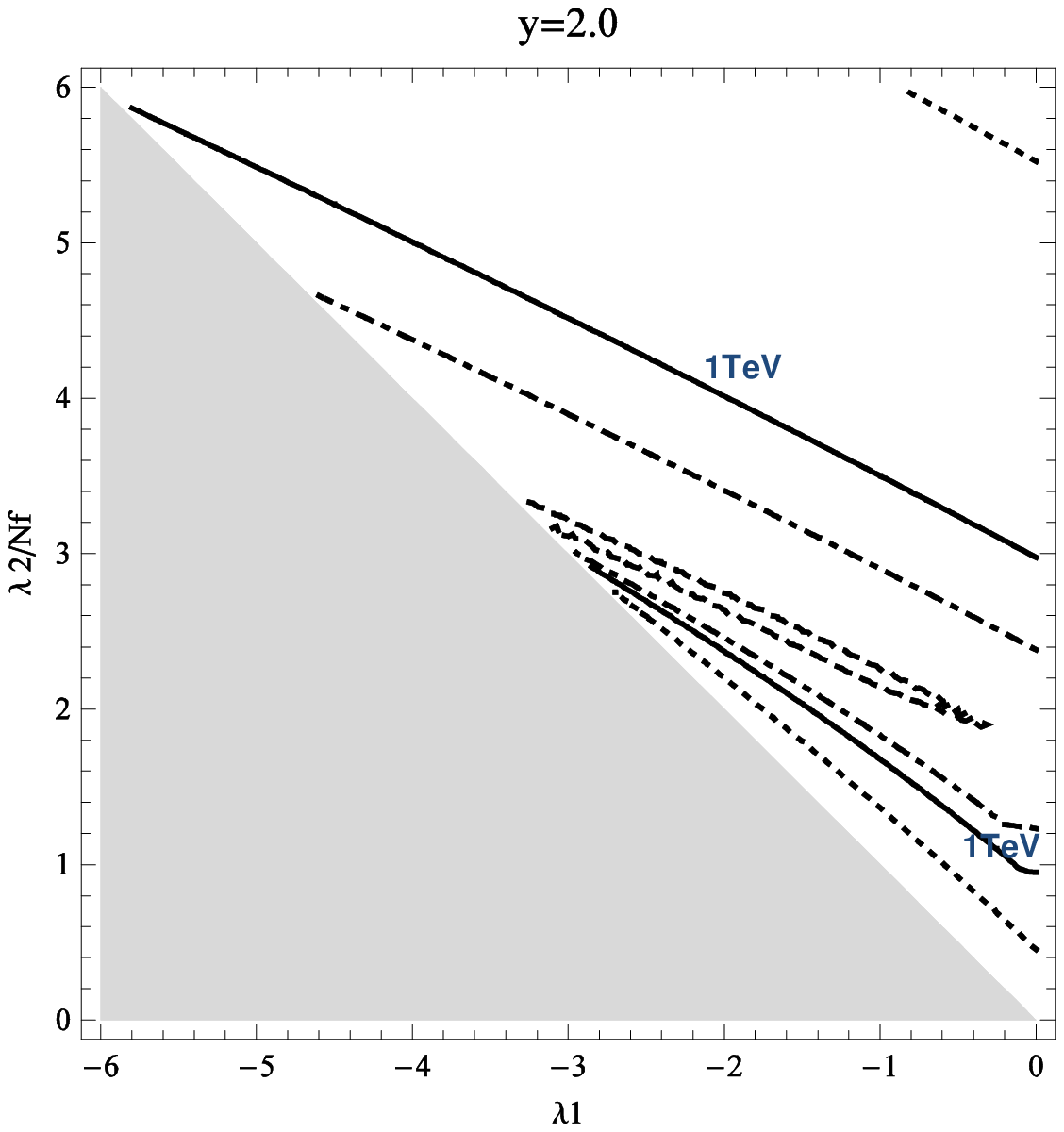}
\caption{
An estimate of the cutoff scale $\Lambda$ for $y=2.0$ in $\lambda_1$--$\lambda_2/N_{f}$ plane.
The dashed, dot-dashed, solid and dotted contours correspond to $\Lambda=5.0, 1.5, 1.0, 0.5$ TeV,  respectively.
In the shaded region, the effective potential at tree-level is unstable and we consider the region 
where $\lambda_1+\lambda_2/N_f\geq0$ only. 
}
\label{cutoffy2}
\vspace{2em}
\includegraphics[width=6cm,clip]{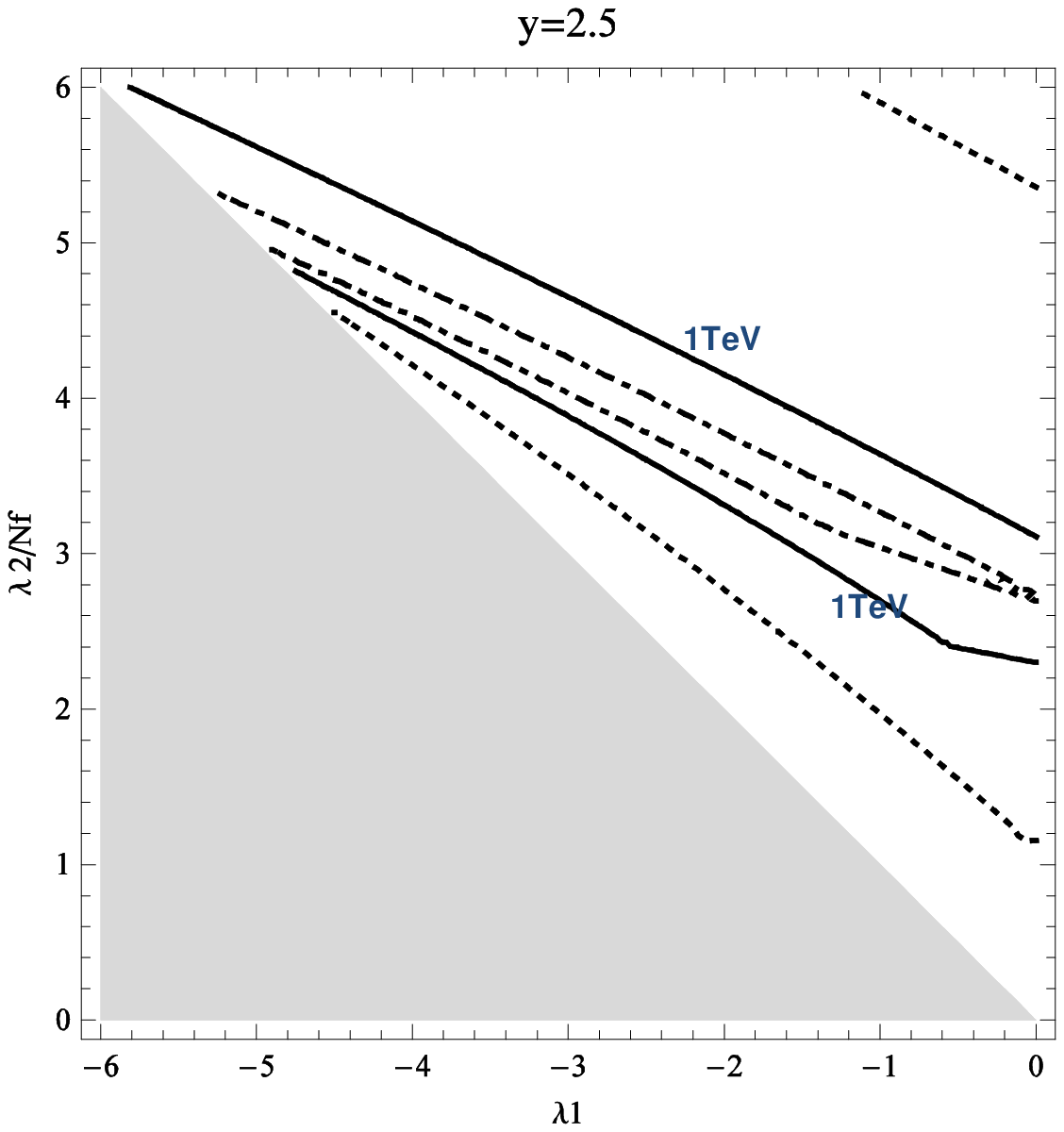}
\caption{
 An estimate of the cutoff scale $\Lambda$ for $y=2.5$ in $\lambda_1$--$\lambda_2/N_{f}$ plane.
 The dot-dashed, solid and dotted contours correspond to $\Lambda=1.5,1.0,0.5$ TeV, respectively.
}
\label{cutoffy2p5}
\end{center}
\end{figure}

\begin{figure}[t]
\begin{center}
\includegraphics[width=6cm,clip]{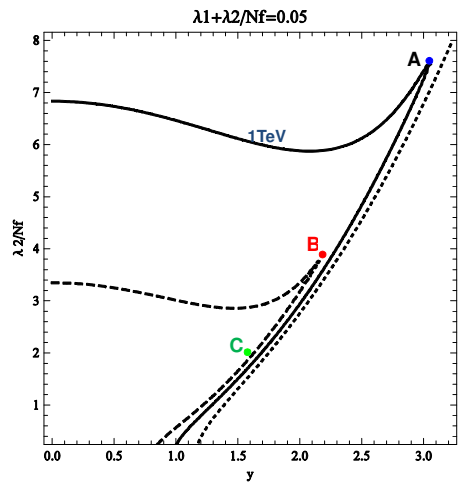}
\caption{
 An estimate of the cutoff $\Lambda$ for $\lambda_1+\lambda_2/N_f=0.05$ in $y$--$\lambda_2/N_{f}$ plane. 
The dashed, solid and dotted contours correspond to $\Lambda=5.0,1.0,0.5$ TeV, respectively.
The blue filled-circle (indicated with "A"), which is  located at the "cusp" on the boundary of the allowed region with  
$\Lambda \ge 1.0$ TeV, indicates the electroweak-scale values of the running coupling constants 
subject to the compositeness condition (Criterion A) at $\Lambda_{\rm 4f}=1.0$ TeV. 
}
\label{cutoff_lam0p05A}
\end{center}
\end{figure}

In FIG.~\ref{cutoffy2}, we show the contours of the estimated cutoff $\Lambda$   for $y=2.0$
in $\lambda_1$--$\lambda_2/N_{f}$ plane.
We see that in the most region, the cutoff scale is around 1 TeV or lower.
For the fixed $\lambda_1$, $\Lambda$ tends to increase with  $\lambda_2/N_f$ for small $\lambda_2/N_f$ 
and tends to decrease with  $\lambda_2/N_f$ for large $\lambda_2/N_f$.
The former (latter) behavior is due to the fact that $\Lambda$ is determined via the vacuum instability (the Landau pole) conditions in that region.
FIG.~\ref{cutoffy2p5} is the similar plot for $y=2.5$.
We see that for the larger value of the Yukawa coupling, the relatively larger values of $\lambda_2/N_f$
are required to fulfill the vacuum stability condition. 
In FIG.~\ref{cutoff_lam0p05A}, we show the contours of the estimated cutoff $\Lambda$ for  $\lambda_1+\lambda_2/N_f=0.05$ in $y$--$\lambda_2/N_{f}$ plane.



In order to ensure the applicability of the effective renormalizable theory, 
the cutoff $\Lambda$ should be taken to be large enough compared 
with other mass scales in the theory:  $\Lambda \gg m_i(\phi_0), \phi_0$.
In the following analysis of the first-order EWPT, it turns out that 
the largest mass scale is given by $m_\xi(\phi_0)$ around 400--700 GeV. 
Then, we require $\Lambda \ge 1$ TeV and exclude the region of the parameter space 
where this condition is not fulfilled. 
This requirement leads  to the constraint on the quark masses, $m_{q^\prime}\lesssim 370$ GeV at tree-level, corresponding to  the Yukawa coupling $y\leq3.0$.


\subsection{Compositeness condition}

Just below the scale of the four-fermion interaction $\mu \lesssim \Lambda_{\rm 4f}$, 
the four-fermion theory  Eq.~(\ref{eq:four-fermion-theory-equivalent}) 
with only quark fields $q^\prime$, 
\begin{align}
\mathcal{L}_{\rm 4f}^\prime=
\bar{q^\prime}i\Slash{\partial} q^\prime -y_{q^\prime 0}(\bar{q^\prime}_{L} \Phi q_{R}^\prime + c.c.) 
-m_{\Phi 0}^2\tr(\Phi^\dagger\Phi) , 
\end{align}
is renormalized to the Yukawa theory Eq.~(\ref{eq:lagrangian}), where  the renormalized couplings are given by 
\begin{eqnarray}
\bar \lambda_1(\mu) &=&  0 ,   \\
\bar \lambda_2(\mu) &=& \frac{32\pi^2}{N_c} \frac{1}{\ln(\Lambda_{\rm 4f}^2/\mu^2)} ,  \\
\bar y(\mu)^2 &=& \frac{16\pi^2}{N_c} \frac{1}{\ln(\Lambda_{\rm 4f}^2/\mu^2)}.
\end{eqnarray}
(See appendix~\ref{sec:appendix2} for detail.)
In the limit $\mu\to\Lambda_{\rm 4f}$,  one finds
\begin{eqnarray}
\bar \lambda_1(\mu) &\to&  0 ,   \nonumber\\
\label{eq:compositeness-condition-infty}
\bar \lambda_2(\mu) &\to& \infty, \\
\bar y(\mu)^2 &\to& \infty , \qquad \bar \lambda_2(\mu) / \bar y(\mu)^2 \to 2.  \nonumber
\end{eqnarray}
This provides the compositeness condition in terms of the renormalized couplings 
as the 
boundary condition of the renormalization group equations at 
$\mu=\Lambda_{\rm 4f}$ \cite{Bardeen:1989ds}. 

In our effective theory formulated as above,  however, the compositeness condition should be modified.
One can not impose the above condition Eq.~(\ref{eq:compositeness-condition-infty})
literally because 
the values of the couplings $\bar \lambda_2(\mu)$ and $\bar y(\mu)$  must 
exceed the perturbativity bounds. 
But,  
the divergence of $\bar \lambda_2(\mu)$ and $\bar y(\mu)$ is due to the Landau pole. Then,
it seems reasonable in our case to substitute the upper limit of the perturbativity bounds for the compositeness condition:
\begin{eqnarray}
\label{eq:compositeness-condition-A}
&& \bar{\lambda}_1(\Lambda_{\rm 4f}) = 0, \quad
\bar{\lambda}_2(\Lambda_{\rm 4f}) = \frac{16\pi^2}{N_f}, \quad
\bar{y}(\Lambda_{\rm 4f})^2 = \frac{16\pi^2}{N_c},  \nonumber\\
&&\bar \lambda_2(\Lambda_{\rm 4f}) / \bar y(\Lambda_{\rm 4f})^2 
= N_c / N_f \quad\qquad [\text{Criterion A}].
\end{eqnarray}

\begin{figure}[t]
\begin{center}
\includegraphics[width=5.5cm,clip]{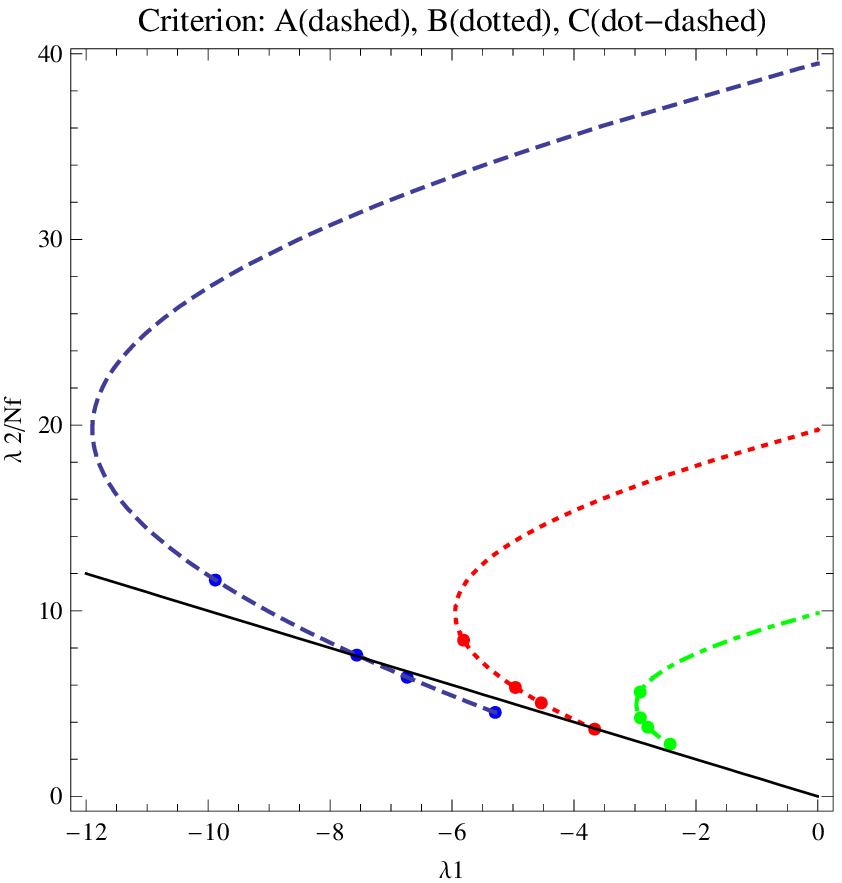}

\vspace{2em}
\includegraphics[width=5.5cm,clip]{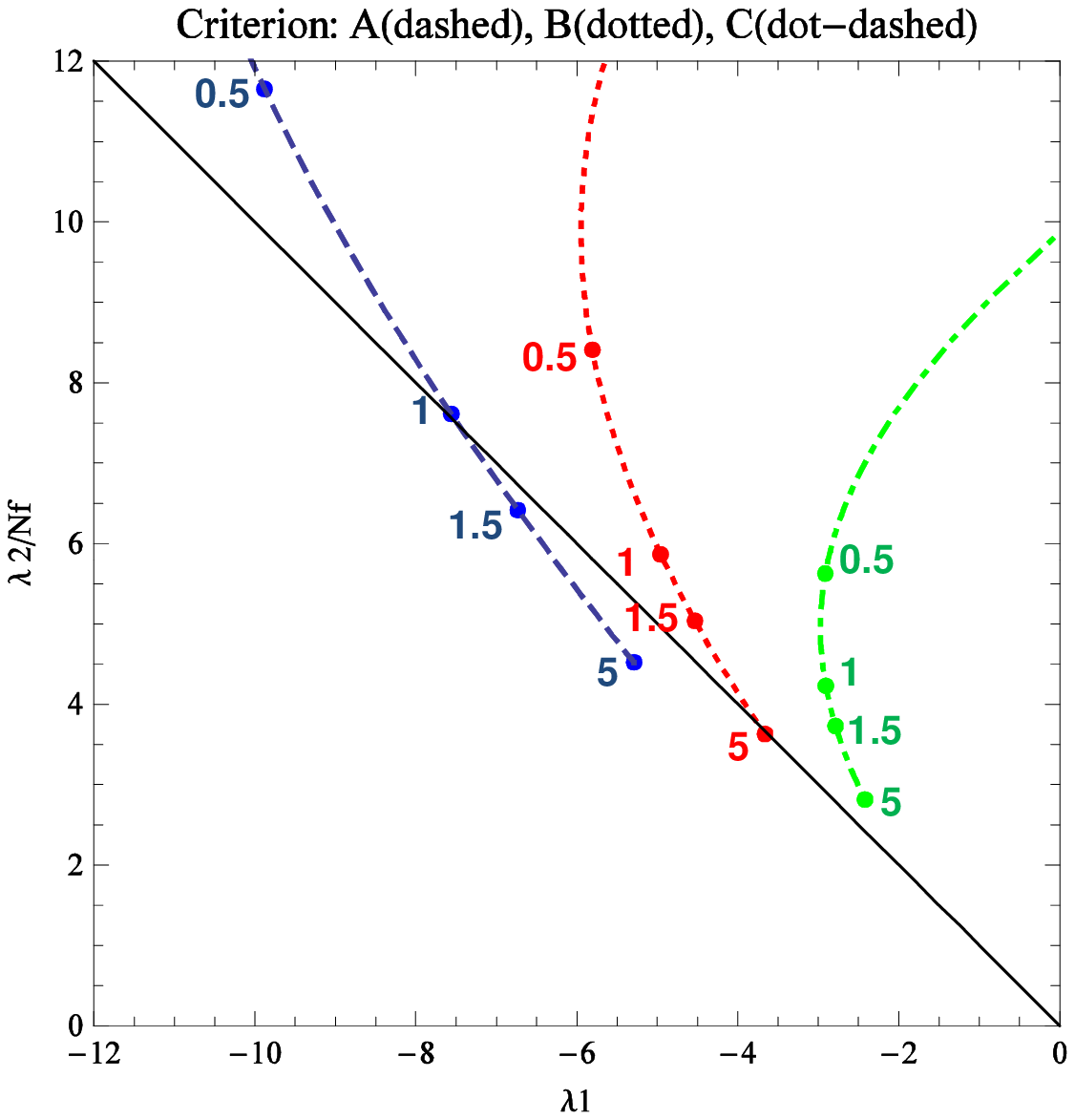}

\caption{ 
The electroweak-scale values of the running coupling constants, which are 
subject to the compositeness conditions at various $\Lambda_{\rm 4f}$, are shown 
in $\lambda_1$--$\lambda_2/N_{f}$ plane.  The blue, red, green data sets correspond to 
the criterion A,  B, C, respectively. 
The blue filled-circles indicate the values $\lambda_1, \lambda_2/N_f$ for $\Lambda_{\rm 4f}$ ($y$) equal to  
0.5 TeV (4.0),  1.0 TeV (3.0),  1.5 TeV (2.7), 5.0 TeV (2.2) 
from top left to bottom right,  respectively.
}
\label{RGflow}
\end{center}
\end{figure}

In FIG.~\ref{RGflow}, we plot the electroweak-scale values $\lambda_1$, $\lambda_2$ of the running coupling constants which are 
subject to the compositeness conditions Eqs.~(\ref{eq:compositeness-condition-A})
at various scales. 
The blue filled-circles correspond to the values of $\Lambda_{\rm 4f}$ ($y$) =
0.5 TeV (4.0),  1.0 TeV (3.0),  1.5 TeV (2.7), 5.0 TeV (2.2) 
from top left to bottom right,  respectively.
The case with $\Lambda_{\rm 4f}=1.0$ TeV comes close to the stability boundary at the electroweak scale, 
taking the value $\lambda_1 + \lambda_2/N_f = 0.05$. 
Then, in FIG.~\ref{cutoff_lam0p05A}, the values $y$, $\lambda_2$ of this case are indicated 
by the blue filled-circle, which is  located at the "cusp" on the boundary of the allowed region with  
$\Lambda \ge 1.0$ TeV. 
%

In fact, the electroweak-scale values of the couplings are rather sensitive to the choice of their
values at the compositeness scale $\Lambda_{\rm 4f}$.  To see this, 
let us refer the above criterion for the compositeness condition as "A", and introduce slightly modified 
criteria "B" and "C" as follows:
\begin{eqnarray}
\label{eq:compositeness-condition-B}
&& \bar{\lambda}_1(\Lambda_{\rm 4f}) = 0, \quad
\bar{\lambda}_2(\Lambda_{\rm 4f}) = \frac{8\pi^2}{N_f}, \quad
\bar{y}(\Lambda_{\rm 4f})^2 = \frac{8\pi^2}{N_c},  \nonumber\\
&&\bar \lambda_2(\Lambda_{\rm 4f}) / \bar y(\Lambda_{\rm 4f})^2 
     = N_c / N_f \quad\qquad [\text{Criterion B}],
\end{eqnarray}
and 
\begin{eqnarray}
\label{eq:compositeness-condition-C}
&& \bar{\lambda}_1(\Lambda_{\rm 4f}) = 0, \quad
\bar{\lambda}_2(\Lambda_{\rm 4f}) = \frac{4\pi^2}{N_f}, \quad
\bar{y}(\Lambda_{\rm 4f})^2 = \frac{4\pi^2}{N_c},  \nonumber\\
&&\bar \lambda_2(\Lambda_{\rm 4f}) / \bar y(\Lambda_{\rm 4f})^2 
= N_c / N_f \quad\qquad [\text{Criterion C}].
\end{eqnarray}
In FIG.~\ref{RGflow}, the red filled-circles and green filled-circles shows 
the electroweak-scale values $\lambda_1$, $\lambda_2$ of the running coupling constants 
subject to the compositeness conditions Eqs.~(\ref{eq:compositeness-condition-B}) and 
(\ref{eq:compositeness-condition-C}), respectively. 
We will discuss this point further in section~\ref{sec:numerical-results} 
in relation to the analysis of EWPT.
%
%
%

\section{Effective potential}
\label{sec:effective-potential-at-one-loop}

\subsection{Zero-temperature effective potential}

At zero temperature, the one-loop effective potential is given by
\begin{align}
V^{(0)}(\phi) =&V_0(\phi) +V_1^{(0)}(\phi), 
\end{align}
where $V_0$ is the tree-level effective potential, $V_1^{(0)}$ is the one-loop contributions at zero temperature.

 $V_0$,  the tree-level effective potential, is given by
\begin{align}
V_0(\phi)=\frac{1}{2}(m_\Phi^2-c)\phi^2+\frac{1}{8}\left( \lambda_1+\frac{\lambda_2}{N_f} \right)\phi^4 . 
\end{align}

$V_1^{(0)}$, the one-loop contribution at zero temperature, is given by 
\begin{align}
V_1^{(0)}(\phi) =&\frac{1}{64\pi^2}\sum_{i=h,\xi,\eta,\pi,q^\prime}n_im_i^4(\phi)
             \left[\ln\frac{m_i^2(\phi)}{\mu^2}-\frac{3}{2}\right]+\frac{1}{2}A\phi^2 
\label{eq:1loop0} .
\end{align}
$m_i(\phi)$ and $n_i$ are the effective  masses depending on $\phi$ and the number of degrees of freedom, respectively, 
which are  given in Table \ref{tab:lsmm}. 
In the calculation of the loop integral in $V_1^{(0)}$, we have taken the limit $\Lambda\to\infty$ and 
have used the $\overline{\rm MS}$ scheme with a slight modification to renormalize the ultraviolet divergences.
The first term is the one-loop contribution in ordinary $\overline{\rm MS}$ scheme with the renormalization 
scale $\mu$. 
The modification is the existence of the second term which are added to preserve the tree-level VEV 
$\phi_0=\sqrt{-2(m_\Phi^2-c)/(\lambda_1+\lambda_2/N_f)}$ and, then, we set $\phi_0=v$(=246GeV).
The parameter $A$ is determined through 
$0=\partial V_1^{(0)}(\phi)/\partial \phi$ at $\phi=v$ and is given by
\begin{align}
A= -\frac{1}{16\pi^2}\sum_{i=h,\xi,\eta,q^\prime} n_ia_im_i^2(v)\left(\ln\frac{m_i^2(v)}{\mu^2}-1\right).
\end{align}

At one-loop, the Higgs boson mass $m_h$ is shifted from the tree-level value $(m_h)_{tree}=\sqrt{\lambda_1+\lambda_2/N_f}v$.
In this paper, we adopt the following definition for the Higgs boson mass $m_h$ at one-loop:   
\begin{align}
\label{eq:mass-h}
m_h^2&\equiv
\left(\lambda_1+\frac{\lambda_2}{N_f}\right)v^2
+\frac{v^2}{8\pi^2}\sum_{i=h,\xi,q^\prime} n_i a_i^2\ln\frac{m_i^2(v)}{\mu^2}. 
\end{align}
This is the curvature of the effective potential at $\phi=v$ : $V^{(0)\prime\prime}(\phi=v)$ with neglecting the 
contributions from the light or massless scalar bosons $\eta$ and $\pi$s 
\footnote{
The contributions of $\eta$ and $\pi$s are divergent and
may be regarded as artifacts due to the fact that 
$V^{(0)\prime\prime}(\phi=v)$ corresponds to 
the off-shell Higgs boson mass at the zero momentum $p^2=0$, while
the on-shell (physical) mass 
should be finite. 
%
%
To neglect  the contributions from $\eta$ and $\pi$s are expected to be valid as long as 
$\lambda_1+\lambda_2/N_f$ is much smaller than $\lambda_2/N_f$ or $y$. 
This argument is in line with \cite{Anderson:1991zb}.
}.

As for the mass of the extra scalar bosons $\xi$, $\eta$ and the fourth-family quarks $q^\prime$, 
we adopt the formula at the tree-level:
\begin{equation}
\label{eq:mass-xi}
m_\xi^2 \equiv 2c+\frac{\lambda_2}{N_f} v^2~,~m_\eta^2\equiv 2c~,~m_{q^\prime}^2\equiv \frac{y^2}{2N_f}v^2 . 
\end{equation}
In the following analysis, we use these definitions for the masses.

\subsection{Finite-temperature effective potential}
 
The one-loop contribution at finite temperature, $V_1^{(T)}$, is given by 
\begin{align}
&V_1^{(T)}(\phi,T) \notag \\
&=\frac{T^4}{2\pi^2}\left(\sum_{i=h,\xi,\eta,\pi}n_iJ_B[m_i^2(\phi)/T^2]+n_{q^\prime} J_F[m_{q^\prime}^2(\phi)/T^2]\right),
\end{align} 
where $J_B$ and $J_F$ are defined by  
\begin{align}
J_B(a)&=\int_{0}^{\infty}dx~x^2 \ln\left(1-e^{-\sqrt{x^2+a}}\right), \notag\\
J_F(a)&=\int_{0}^{\infty}dx~x^2 \ln\left(1+e^{-\sqrt{x^2+a}}\right).
\end{align}
In the following analysis of the EWPT, we carry out a numerical integral for $J_B$ and $J_F$ without high temperature expansion. 


In the ordinary perturbation theory at finite temperature, the perturbative expansion breaks down near the critical 
temperature due to the existence of the higher-loop IR divergent diagrams in the massless limit.
To improve the reliability of the perturbative expansion, we include the contributions from the ring diagrams 
which are the most dominant IR contributions at each order of the perturbative expansion
\cite{Fendley:1987ef,Espinosa:1992gq,Parwani:1991gq, Carrington:1991hz, Dine:1992wr,Arnold:1992rz,Fodor:1994bs}
\footnote{
 The ring-improved perturbation theory is valid when the non-ring diagrams 
 are suppressed with respect to the ring-diagrams.
 We infer the expansion parameter in the ring-improved perturbation theory by the power counting argument 
 as in the $\lambda\phi^4$ theory \cite{Fendley:1987ef}. 
 By inspecting the higher order diagrams for the scalar field self-energies,  
 the expansion parameters are expected to be 
 $n_i \, \text{max}(\lambda_1+\lambda_2/N_f,\lambda_2/N_f)T/[4\pi \mathcal{M}_{i}(\phi,T)]$ 
 ($i=h, \xi, \eta, \pi$), 
 where $\mathcal{M}_{i}(\phi,T)$ and $n_i$ are the effective T-dependent masses of the scalar bosons and the corresponding 
 numbers of degrees of freedom.  
 This is maximized for $i=\pi$ and $\phi=0$, then, 
 $(N_f^2-1) \text{max}(\lambda_1+\lambda_2/N_f,\lambda_2/N_f)T_c/[4\pi \mathcal{M}_{\pi}(0,T_c)\ll 1$ 
 should be held for the ring-improved perturbation theory to be valid at $T=T_c$.
 In the analysis of the EWPT, we have observed that for $y\gtrsim 2.1$ the above expansion parameter is 
 smaller than 1 but is not so small (greater than 0.5).
}.

One can include the contribution of ring diagrams,  $V_{\ring}(\phi,T)$,  by replacing $m_i^2(\phi)$ ($i=h,\xi,\eta,\pi$) in $V_1^{(0)}$ and $V_1^{(T)}$ with the effective T-dependent masses $\mathcal{M}_i^2(\phi,T)\equiv m_i^2(\phi)+\Pi_\Phi$,
where $\Pi_\Phi$ is the self-energy of the scalar bosons in the IR limit where the Matsubara frequency and the momentum of the external fields go to zero and in the leading order of $m_i(\phi)/T$ and is given by
\begin{align}
\Pi_\Phi =\frac{1}{12}[(N_f^2+1)\lambda_1+2N_f\lambda_2+N_c y^2]T^2,
\end{align}
at one-loop order.

After all, the one-loop ring-improved effective potential is given by
\begin{widetext}
\begin{align}
V(\phi,T) =&V_0(\phi)+V_1^{(0)}(\phi) +V_1^{(T)}(\phi,T)+ V_{\ring}(\phi,T)  \notag \\
        =&V_0(\phi)+  \frac{1}{2}A\phi^2 
         +\sum_{i=h,\xi,\eta,\pi}n_i\left[\frac{1}{64\pi^2}\mathcal{M}_i^4(\phi,T)
          \left(\ln\frac{\mathcal{M}_i^2(\phi,T)}{\mu^2}-\frac{3}{2}\right)
          +\frac{T^4}{2\pi^2}J_B[\mathcal{M}_i^2(\phi,T)\beta^2]\right] \notag \\
         &+n_{q^\prime}\left[\frac{1}{64\pi^2}m_{q^\prime}^4(\phi)
             \left(\ln\frac{m_{q^\prime}^2(\phi)}{\mu^2}-\frac{3}{2}\right)
                +\frac{T^4}{2\pi^2}J_F[m_{q^\prime}^2(\phi)\beta^2]\right]. 
\label{eq:ftep}                
\end{align}
\end{widetext}
In the following, we study the finite-temperature EWPT by numerically 
evaluating this effective potential.

\section{Numerical analysis of electroweak phase transition}
\label{sec:numerical-results}

The EWPT should be strongly first order in order to avoid the washout of the generated baryon asymmetry in the broken phase. 
How strongly first order  the phase transition must be depends on the energy of the sphaleron 
solution \cite{Manton:1983nd,Klinkhamer:1984di}
in the model considered.  As long as the classical (static) solution of the equation of motions is concerned, 
one may neglect the effect of the fourth-family quarks even when they are heavy \footnote{The large Yukawa couplings of the fourth-family quarks and leptons may affect the baryon number diffusion rate, which is usually computed at one-loop 
including the effect of the fluctuations around the sphaleron 
solution \cite{Arnold:1987mh,Akiba:1989xu,Carson:1990jm,Baacke:1993aj,Baacke:1994ix,
Moore:1995jv,Kovner:1999ja}.}. 
Then, one may use the condition 
\begin{equation}
\phi_c/T_c \, \gtrsim \, 1,  
\end{equation}
as the criterion for a strongly first-order EWPT, 
 as discussed in our previous work \cite{Kikukawa:2007zk}. 


In the following analysis, assuming a first-order phase transition, we solve the conditions 
\begin{align}
\frac{\partial V(\phi, T_c)}{\partial\phi}\bigg|_{\phi=\phi_c}=0, ~~V(\phi_c,T_c)=V(0,T_c),
\label{eq:condition-first-order}
\end{align}
numerically for various parameters,  $m_h$, $m_\xi$, $m_{q^\prime}$ and 
$m_\eta$  (see Eqs.~(\ref{eq:mass-h}), (\ref{eq:mass-xi}) for definitions). Then we evaluate 
$\phi_c/T_c$ in order to estimate the strength of the first-order phase transition.  

We first explore the parameter space of the renormalizable theory without the constraint due to the compositeness condition, 
in order to locate  the parameter region where a strongly first-order EWPT is realized. 
We consider  only the region 
where the stability condition at tree-level, $\lambda_1+\lambda_2/N_f\geq 0$, is satisfied, 
otherwise the mass parameter squares in the one-loop effective potential become negative. 
We also consider only the region where the perturbation theory is reliable:
\begin{align}
\lambda_1+\lambda_2/N_f \ll \frac{(4\pi)^2}{N_f^2}, \, \frac{\lambda_2}{N_f} \ll \frac{(4\pi)^2}{N_f^2}, \,
y^2 \ll \frac{(4\pi)^2}{N_c}.
\end{align}
In order to fulfill the requirement $\Lambda\geq 1$ TeV, we also concentrate on a region where $y\leq 3.0$.  
For an estimated $\Lambda$, we require that $\phi=v$ is the global minimum of the one-loop zero-temperature 
effective potential $V^{(0)}(\phi)$ for $0<\phi<\Lambda$.
The renormalization scale of the effective potential is set to the electroweak scale, $\mu=v$.
We next examine the possible overlap of the parameter region of the strongly first-order EWPT and
the region subject to the compositeness condition.


\subsection{Yukawa theory without compositeness condition 
}

\begin{figure}[t]
\begin{center}
\includegraphics[width=6cm,clip]{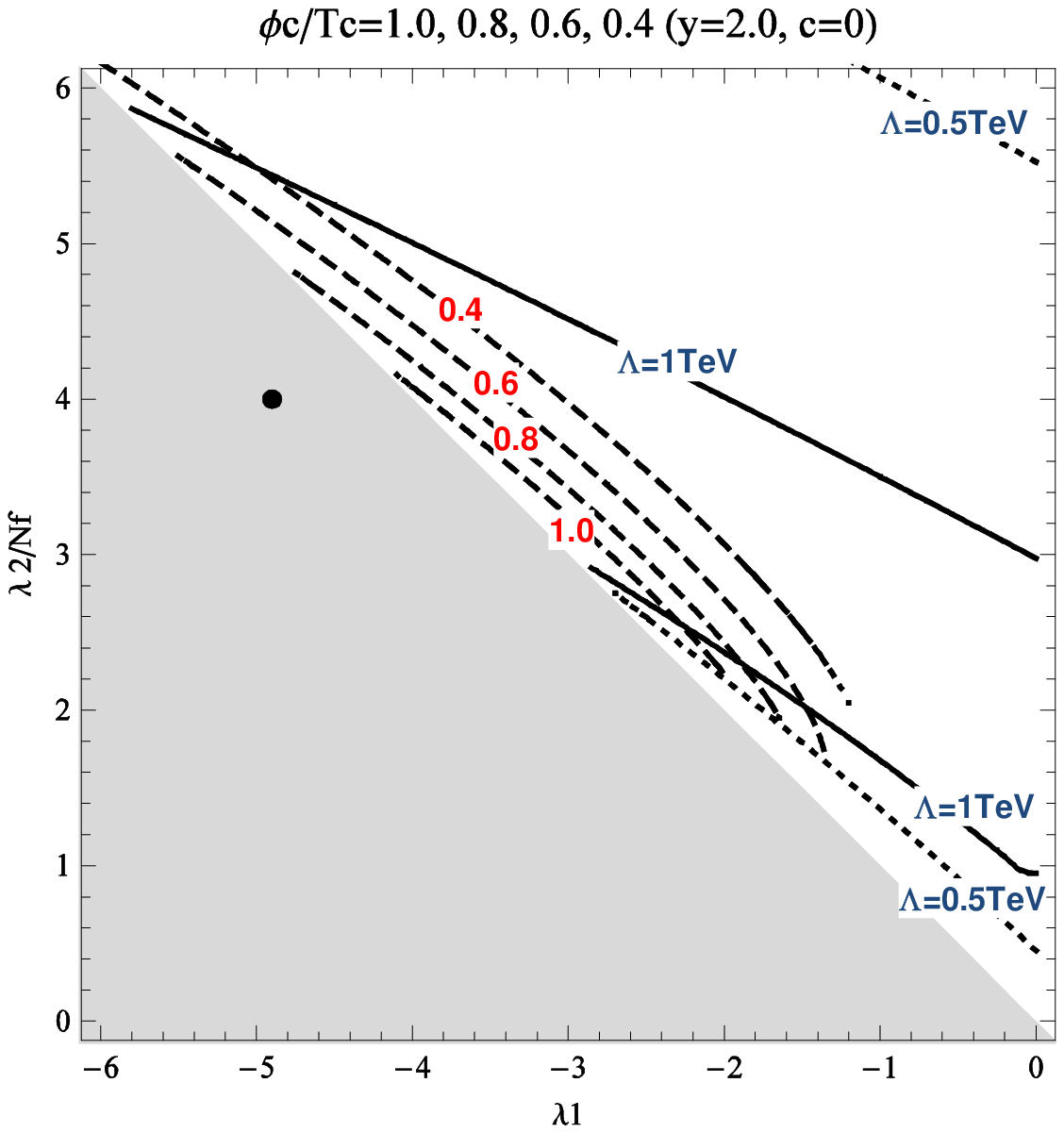}
\caption{ 
Contour plots for various $\phi_c/T_c$ in $\lambda_1$--$\lambda_2/N_f$ plane at $y=2.0$, $c=0$.
The dashed lines are  the contours of $\phi_c/T_c=1.0, 0.8, 0.6, 0.4$ from bottom left to top right as indicated.
The solid lines indicate the boundary of the allowed region with $\Lambda\geq1$TeV.
(The region with $1$ TeV$>\Lambda\geq0.5$TeV is also shown.)
In the shaded region,  the effective potential at tree-level is unstable.
The black filled-circle indicates the values of the couplings $\lambda_1, \lambda_2/N_f$ when  the compositeness condition is satisfied at $\Lambda_{\rm 4f}$= 9 TeV so that $y=2.0$.
}
\label{PD66y2}
\vspace{2em}
\includegraphics[width=6cm,clip]{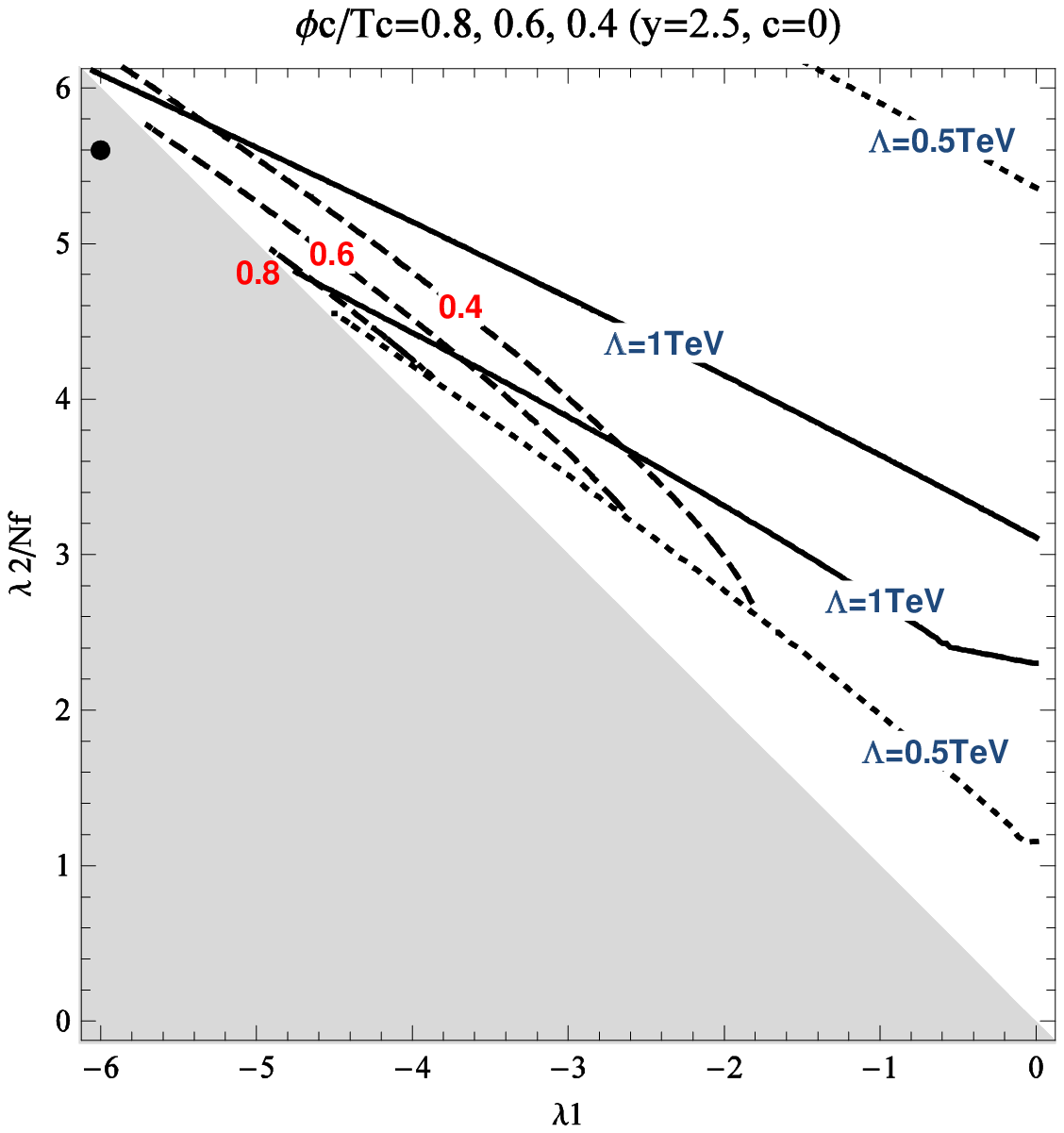}
\caption{ 
Contour plots for various $\phi_c/T_c$ in $\lambda_1$--$\lambda_2/N_f$ plane at $y=2.5$, $c=0$.
The dashed lines are the contours of $\phi_c/T_c=0.8, 0.6, 0.4$ from bottom left to top right as indicated. 
The black filled-circle indicates the values of the couplings $\lambda_1, \lambda_2/N_f$ when  the compositeness condition is satisfied at $\Lambda_{\rm 4f}$= 2.3 TeV so that $y=2.5$.
}
\label{PD66y2p5}
\end{center}
\end{figure}

We first discuss  the numerical results for the Yukawa theory 
without the constraint due to the compositeness condition.
In this analysis, we neglect the effect of the U(1)$_A$ symmetry breaking term, setting $c=0$.  

In FIG.~\ref{PD66y2}, we show the contours of $\phi_c/T_c$
in the $\lambda_1$--$\lambda_2/N_f$ plane for $y=2.0$ in 
the allowed region with $\Lambda\geq 1$ TeV.
(The contours with 1 TeV$\geq\Lambda \geq $0.5 TeV are also shown for reference.)
We clearly see that the region of the strongly first-order transition lies above the stability boundary $\lambda_1+\lambda_2/N_f = 0$.  
When $\lambda_1+\lambda_2/N_f$ gets larger, $\phi_c/T_c$ tends to become smaller.
For a fixed small $\lambda_1+\lambda_2/N_f$, there is a tendency that 
$\phi_c/T_c$ decreases as $\lambda_2/N_f$ increases. 

FIG.~\ref{PD66y2p5} is a similar contour diagram for $y=2.5$.  
We notice that the region with $\phi_c/T_c>1$ disappears in this case. 
The substantial region close to the stability boundary is excluded by the condition 
$\Lambda\geq 1$ TeV.  This is because one encounters the instability at a lower scale for the larger Yukawa coupling. But, even when the allowed region is extended to 
$\Lambda\geq 0.5$ TeV, the transition is weakly first order with  $\phi_c/T_c < 1$. 
For larger values of $y$,  as far as we explored, 
the transition gets more weakly first order, or possibly second order.

\begin{figure}[b]
\begin{center}
\includegraphics[width=6cm,clip]{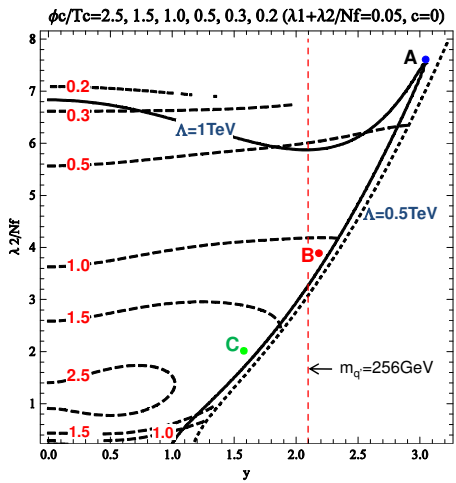}
\caption{ 
Contour plots for various $\phi_c/T_c$ in $y$--$\lambda_2/N_f$ plane at $\lambda_1+\lambda_2/N_f=0.05$, $c=0$.
The value of $\phi_c/T_c$ for each contour is shown in the diagram. 
}
\label{PD48lam0p05}
\end{center}
\end{figure}

In FIG.~\ref{PD48lam0p05}, we show the contours of $\phi_c/T_c$
in $y$--$\lambda_2/N_{f}$ plane for $\lambda_1+\lambda_2/N_f=0.05$.
We  see clearly that 
a strongly first-order phase transition is realized even
for $y \gtrsim 2.1$, which corresponds to the experimental lower bound 
from the direct search of the fourth-family quarks $m_{q^\prime}\gtrsim 256$ GeV,  
in the range $3\lesssim \lambda_2/N_f\lesssim 4$ (corresponding to 430 GeV$\lesssim m_\xi\lesssim$500 GeV).
For larger  values of $y$, however,  the stability of the electroweak vacuum  requires lager 
values of $\lambda_2/N_f$ and then the strength of the first-order phase transition becomes weaker, 
or possibly second order. 



\begin{figure}[t]
\begin{center}
\includegraphics[width=5.5cm,clip]{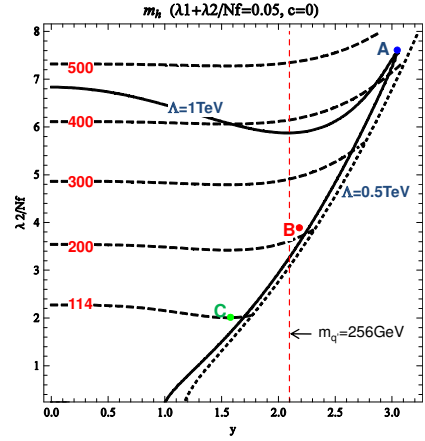}
\caption{ 
Contour plots for various $m_h$ in $y$--$\lambda_2/N_f$ plane at $\lambda_1+\lambda_2/N_f=0.05$, $c=0$.
The dashed lines are the contours of $m_h=114, 200, 300, 400, 500$ GeV as indicated. 
}
\label{mh-plot}
\end{center}
\end{figure}

In FIG.~\ref{mh-plot}, we show the contours of the one-loop Higgs mass $m_h$ 
in $y$--$\lambda_2/N_{f}$ plane for $\lambda_1+\lambda_2/N_f=0.05$, corresponding to $(m_h)_{\rm tree}=55$ GeV.
In this case, for $\lambda_2/N_f\gtrsim 1$ or $y\gtrsim 1$ we can neglect the one-loop contributions 
of $\eta$ and $\pi$s to $m_h$ and the use of the formula Eq.~(\ref{eq:mass-h}) is validated. 
It is clear that for $y \gtrsim 2.1$ the Higgs mass receives rather large one-loop correction and 
is heavier than experimental lower bound, $m_h>114$GeV, in the whole region.

\begin{figure}[t]
\begin{center}
\includegraphics[width=5.5cm,clip]{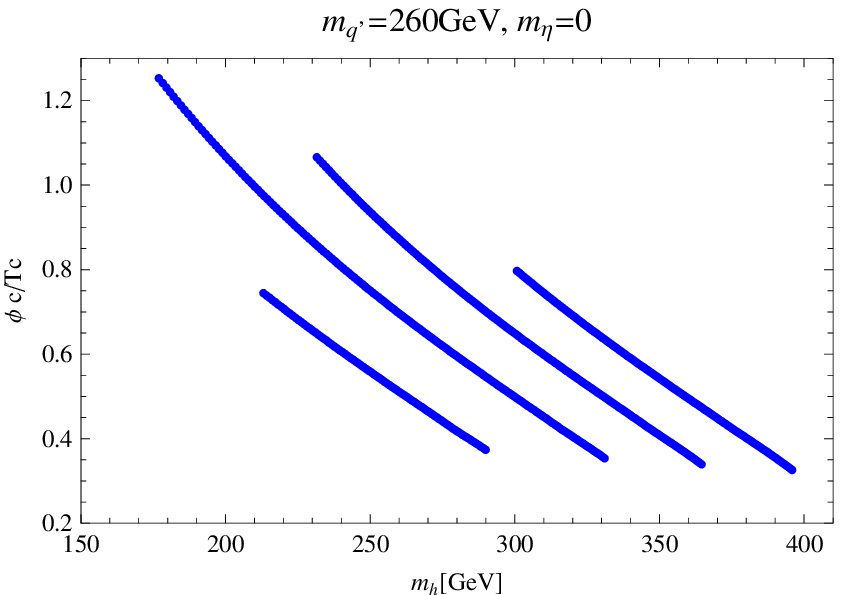}
\caption{ 
$\phi_c/T_c$ as a function of $m_h$ for $m_{q^\prime}=260$ GeV and $m_\xi=400, 450, 500, 550$ GeV from bottom left to top right. 
}
\label{changing_mxi}
\vspace{2em}
\includegraphics[width=5.5cm,clip]{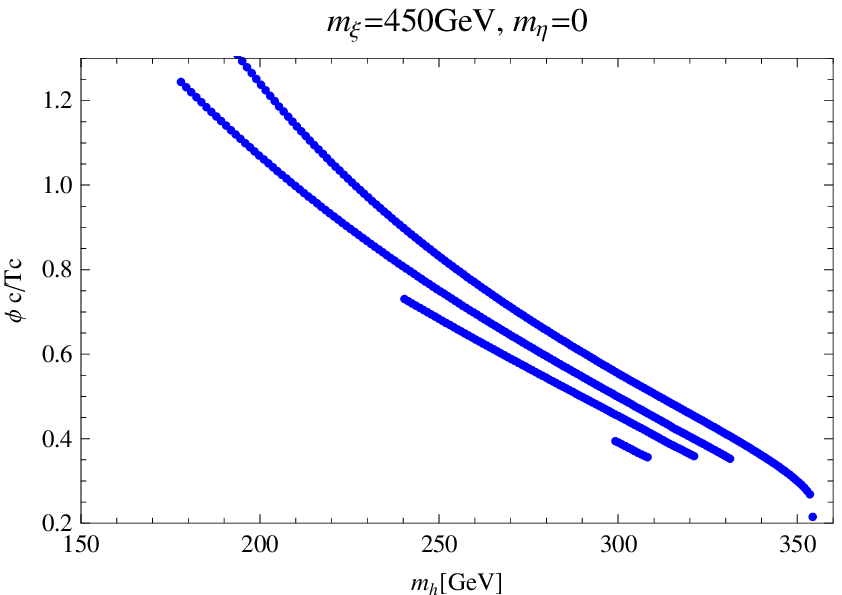}
\caption{ 
$\phi_c/T_c$ as a function of $m_h$ for $m_\xi=450$ GeV and $m_{q^\prime}=200, 260, 280, 300$ GeV from top to bottom. 
}
\label{changing_mf}
\end{center}
\end{figure}

In FIG.~\ref{changing_mxi},  $\phi_c/T_c$ is plotted as a function of $m_h$ for several values of $m_\xi$ 
with $m_{q^\prime}=260$ GeV fixed.
We can see that with $m_\xi$ fixed, $\phi_c/T_c$ decreases as $m_h$ increases, 
while, with $m_h$ fixed, $\phi_c/T_c$ increases as $m_\xi$ increases.
In the whole range, $m_h$ exceeds the experimental bound, $m_h > 114$GeV. 
In FIG.~\ref{changing_mf}, on the other hand, 
$\phi_c/T_c$ is plotted as a function of $m_h$ for several values of $m_{q^\prime}$ with $m_\xi=450$GeV fixed.
In this case, we note that with $m_h$ fixed, $\phi_c/T_c$ decrease as $m_{q^\prime}$ increases.

\subsection{Effect of explicit symmetry breaking term}

\begin{figure}[h]
\begin{center}
\includegraphics[width=5.5cm,clip]{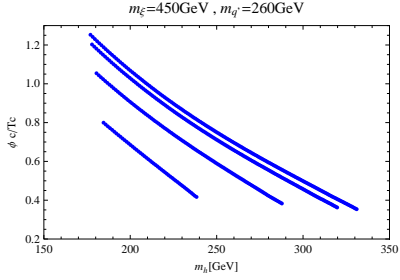}
\caption{ 
$\phi_c/T_c$ as a function of $m_h$ for $m_\xi=450$ GeV, $m_{q^\prime}=260$ GeV 
and $m_\eta=0, 50, 100, 150$ GeV from top to bottom. 
}
\label{changing_meta}
\end{center}
\end{figure}

We next examine the effect of the explicit U(1)$_A$ symmetry breaking term 
by taking a value of $m_\eta$ (or $c$) non-zero. 
The effect can be read from FIG.~\ref{changing_meta}.
For a fixed value of $m_h$, $\phi_c/T_c$ decreases as $m_\eta$ increases. 
So we see that the explicit symmetry breaking term reduces the strength of the first-order phase transition.
The experimental bound on the mass of the pseudo NG boson depends on how $\eta$ couples to the other particles, 
which is not specified in our model. 
If we adopt the bound for the pseudoscalar Higgs boson in supersymmetric models (with $\tan\beta>0.4$), 
the allowed values are $m_\eta > 93.4$ GeV \cite{Amsler:2008zz}.
For such a value of $m_\eta$, we see that the region of $\phi_c/T_c\geq 1$ becomes quite narrow 
\footnote{We note that the tree-level formula is used for $m_\eta$ in this analysis and an inclusion of 
the one-loop correction to $m_\eta$ would affect the extent of the region with $\phi_c/T_c\geq 1$.}. 


\begin{figure}[h]
\begin{center}
\includegraphics[width=6cm,clip]{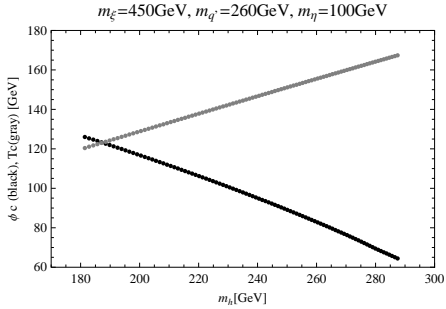}
\caption{ 
Plots of $\phi_c$ (black) and $T_c$ (gray) as a function of $m_h$ with 
$m_{q'}=260$ GeV, $m_\xi=450$ GeV and $m_\eta=100$ GeV fixed. 
}
\label{phic-Tc}
\end{center}
\end{figure}

Typical values of $\phi_c$ and $T_c$ are shown separately in FIG.~\ref{phic-Tc} as 
a function of $m_h$ for 
$m_{q'}=260$ GeV, $m_\xi=450$ GeV and $m_\eta=100$ GeV fixed. 

\subsection{Compositeness condition}

Finally, 
we discuss  the numerical results for the Yukawa theory 
with the compositeness condition imposed. 
Let us recall the plots in  FIG.~\ref{cutoff_lam0p05A} and FIG.~\ref{RGflow} 
of the electroweak-scale values of the running coupling constants 
which are 
subject to the compositeness conditions Eqs.~(\ref{eq:compositeness-condition-A}) (Criterion A)
at various scales: 
the values $\lambda_1$, $\lambda_2$ for  $\Lambda_{\rm 4f}=1.0$ TeV 
come close to the stability boundary at the electroweak scale, 
taking the value $\lambda_1 + \lambda_2/N_f = 0.05$ and 
it is located at 
the "cusp" on the boundary of the allowed region with  
$\Lambda \ge 1$ TeV.  
In this case, corresponding to the blue filled-circle (indicated with "A") in FIG.~\ref{PD48lam0p05}, 
though we could not find the solution of Eqs.~(\ref{eq:condition-first-order}), 
we have checked that the phase transition is weakly first order with $\phi_c/T_c <0.001$, or possibly second order.



The above conclusion deserves discussions. 
In case if one adopts  the other criterion for the compositeness condition, 
the phase transition can be first order.  In the case of the criterion B, for example, 
the electroweak-scale values of the running coupling constants which are 
subject to the compositeness condition at $\Lambda_{\rm 4f}=3.7$ TeV 
come close to the stability boundary at the electroweak scale, 
taking the value $\lambda_1 + \lambda_2/N_f = 0.05$. 
In FIG.~\ref{PD48lam0p05}, we show
the electroweak-scale values $y$, $\lambda_2$ of this case 
by the red filled-circle (indicated with "B"). 
One immediately see that 
the phase transition is rather strongly first order with $\phi_c/T_c \simeq 1$. 
Thus
the critical behavior of the phase transition is rather sensitive to the choice of the
values of the coupling constants at the compositeness scale $\Lambda_{\rm 4f}$. 
However, combining with the results of the Yukawa theory without the compositeness condition, 
in particular, with the result for $y=2.5$ shown in FIG.~\ref{PD66y2p5}, 
it seems fair to say that 
{\em the four-fermion interaction of the fourth-family quarks,  
which causes EWSB at zero-temperature and produces
the mass of the heavy quarks larger than $m_{q'} \simeq 310$ GeV 
($y \gtrsim 2.5$, $\Lambda_{\rm 4f}  \lesssim 2.3$ TeV),   
does not lead to the strongly first-order EWPT at finite temperature} \footnote{There remains a 
possibility that the effect of the electroweak interactions might enhance slightly the strength 
of the first-order phase transition.}. 
We note that a region with $\Lambda_{\rm 4f}< 1$ TeV ($y> 3.0$, $m_{q^\prime}> 370$ GeV) 
is beyond the scope of our analysis using the effective renormalizable theory.



%

\begin{figure}[h]
\begin{center}
\includegraphics[width=6cm,clip]{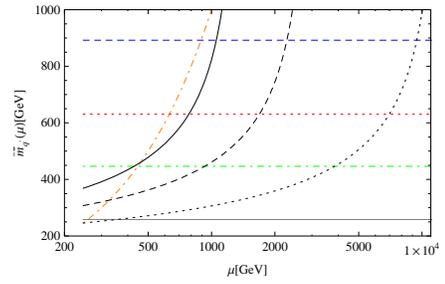}
\caption{ 
Plots of the running mass
of the fourth-family quarks,  $\bar m_{q^\prime}(\mu) = \bar y(\mu) v / \sqrt{2 N_f}$, 
as a function of the renormalization scale $\mu$. 
The value of "on-shell" mass is  defined by 
$m_{q^\prime ,phys} = \bar y(m_{q^\prime ,phys}) v / \sqrt{2 N_f}$.
The dot-dashed line (orange) indicates  $\bar m_{q^\prime}(\mu) = \mu$. 
}
\label{fermion-mass-running}
\end{center}
\end{figure}

\section{Conclusion and Discussion}
\label{sec:conclusion-discussion}

In this paper, we have studied the finite-temperature electroweak phase transition in the model where 
the electroweak symmetry is dynamically broken due to the four-fermion interaction of the fourth-family quarks. 
Based on the effective renormalizable Yukawa theory with/without compositeness condition, 
we have estimated the strength of the first-order phase transition $\phi_c/T_c$, 
using the finite temperature effective potential at one-loop with the ring-improvement.
In the Yukawa theory without compositeness condition, the phase transition can be strongly 
first order with 
$\phi_c/T_c\gtrsim 1$ for the experimentally acceptable Higgs boson and fourth-family quarks masses in the range 
256 GeV $\lesssim m_{q^\prime} \lesssim$ 290 GeV when $\lambda_1+\lambda_2/N_f=0.05$.
On the other hand, once the compositeness condition is imposed as the boundary condition for the 
running coupling constants,  
the values of the couplings at the electroweak scale are restricted 
in a certain region of the parameter space. There, 
the phase transition turns out to be weakly first order, or possibly second order.
The above result depends on how to specify the compositeness condition. 
In fact, we observed that 
the values of the running couplings at $\mu=v$ are rather sensitive to the choice 
of the boundary condition at $\mu=\Lambda_{\rm 4f}$ and if one takes the smaller values as the boundary condition at $\mu=\Lambda_{\rm 4f}$ 
(for example, Criterion B), the phase transition can be strongly first order 
with $\phi_c/T_c\sim \mathcal{O}(1)$.
In spite of this ambiguity, combining with the results of the Yukawa theory without the 
compositeness condition,  
it seems plausible that for $3.0\geq y\gtrsim 2.5$
(corresponding to $1$ TeV $\lesssim\Lambda_{\rm 4f} \lesssim 2.3$ TeV 
and $480$ GeV $\gtrsim m_{q', phys} \gtrsim 330$ GeV "on shell" 
\footnote{
 For $y= 3.0$ ( $m_{q^\prime}= 370$ GeV), the fourth-family quark mass defined by Eq.~(\ref{eq:mass-xi}) 
 receive rather large correction due to the running of the Yukawa coupling. 
 The "on-shell" mass, defined by 
 $m_{q^\prime, phys} = \bar y(m_{q^\prime, phys}) v / \sqrt{2 N_f}$, is $m_{q^\prime, phys}\simeq480$ GeV 
 for $y= 3.0$ and is $m_{q^\prime, phys}\simeq 330$ GeV for $y=2.5$ (See FIG.~\ref{fermion-mass-running}).
}), 
the four-fermion interaction of the fourth-family quarks, which causes EWSB at zero temperature, does not lead to the strongly first-order EWPT at finite temperature. 
We note that a region with $y> 3.0$ is beyond the scope of our analysis 
using the effective renormalizable theory.


Since the critical behavior of first-order phase transition at finite temperature  that we concern, is not universal in general, the result of our analysis would depend on our choice of low energy effective theory, where a certain truncation of fields and operators has to be done \cite{Hasenfratz:1991it}.
Our analysis, therefore, must be semi-quantitative (or qualitative),
showing a possibility to realize the strongly first-order EWPT required for the electroweak baryogenesis.

Fortunately, it is now possible to formulate the model considered in this paper
non-perturbatively on the lattice, preserving SU(2)$\times$U(1) chiral gauge symmetry 
exactly \cite{Neuberger:1997fp,Neuberger:1998wv,Kikukawa:1997qh,Ginsparg:1981bj,
Luscher:1998pqa,Luscher:1998du, Kadoh:2007xb}.  
A direct numerical analysis of EWPT by Monte Carlo simulations 
with the state-of-art technique \cite{Neuberger:1998my, Frommer:1995ik, van den Eshof:2002ms,Cundy:2004pza,Chiu:2002eh,Vranas:2006zk,Fukaya:2006vs,Fukaya:2007fb,Aoki:2008tq}
may shed light on the issues discussed above. Even in the global symmetry limit, where 
SU(2)$\times$U(1) gauge interactions are turned off,  there are several questions worth studying: 
one may study the phase structure and the critical behavior of  chiral symmetry restoration at finite temperature in relation to 
the triviality bounds for the masses 
of the Higgs bosons and the fourth-family quarks and leptons \cite{Gerhold:2007yb,Gerhold:2007gx,Gerhold:2008mb}.

In this paper, we have focused on the effect of the fourth-family quarks.
The analysis of the effect of the fourth-family charged lepton and neutrino will be 
reported in a subsequent paper~\cite{Kikukawa-Kohda:2009xx}.

\begin{acknowledgments}
The authors would like to thank Y.~Okada and E.~Senaha for valuable discussions.
Y.K. is supported in part by Grant-in-Aid for Scientific Research No.~14046207.
\end{acknowledgments}
 
\appendix
\section{Scalar potential}  \label{sec:appendix}

In this appendix, we give scalar potential in the full effective Yukawa theory which includes 
both fourth-family quarks and leptons as well as all scalar fields 
$\Phi$, $H_{\tau^\prime}$ and $\chi^a$ ($a=1,2,3$).  

The most general renormalizable scalar potential which is consistent with symmetry (including the general U(1)$_A$ 
breaking terms) is given by
\begin{align}
V(\Phi, H_{\tau^\prime}, \vec{\chi})= V_2+V_3+V_4
\end{align}
where
\begin{eqnarray}
V_2
&=&m_{\Phi}^2\tr(\Phi^\dagger\Phi)+ m_{H_{\tau^\prime}}^2 H_{\tau^\prime}^\dagger H_{\tau^\prime}
+m_\chi^2 \chi^{a*}\chi^a \nonumber\\
&& -c(\det \Phi +c.c.) ,  \\
V_3&=&a H_{\tau^\prime}^T \epsilon \tau^a\chi^{a*} H_{\tau^\prime}+ c.c. ,
\end{eqnarray}
and
\begin{align}
V_4& \notag\\
=& \sum_{i,j,k,l} \left[ \rho_{ijkl} \tr(\phi_i^\dagger\phi_j) \tr(\phi_k^\dagger\phi_l)
+ \rho_{ijkl}^\prime \tr(\phi_i^\dagger\phi_j\phi_k^\dagger\phi_l) \right] \notag\\
&+\alpha (H_{\tau^\prime}^\dagger H_{\tau^\prime})^2+\beta_1(\chi^{a*}\chi^a)^2+\beta_2(\chi^a\chi^a)(\chi^{b*}\chi^{b*}) \notag\\
&+\sum_{i,j} \left[ \gamma_{ij} \tr(\phi_i^\dagger\phi_j)H_{\tau^\prime}^\dagger H_{\tau^\prime}
+ \gamma_{ij}^\prime H_{\tau^\prime}^\dagger \phi_i\phi_j^\dagger H_{\tau^\prime} \right] \notag\\
&+ \sum_{i,j}\gamma_{ij}^{\prime\prime} \tr(\phi_i^\dagger\phi_j) \chi^{a*}\chi^a +\kappa H_{\tau^\prime}^\dagger H_{\tau^\prime}\chi^{a*}\chi^a
\end{align}
where we have used the following notation
\begin{align}
\phi_1\equiv \Phi~,~\phi_2\equiv \epsilon \Phi^* \epsilon,
\end{align}
and $i, j, k, l=1,2$.

\section{Renormalization group evolution of the four-fermion theory} \label{sec:appendix2}

Just below the scale of the four-fermion interaction $\mu \lesssim \Lambda_{\rm 4f}$, the four-fermion theory equivalent, Eq.~(\ref{eq:four-fermion-theory-equivalent}), 
\begin{align}
\mathcal{L}_{\rm 4f}^\prime=
\bar{q^\prime}i\Slash{\partial} q^\prime -y_{q^\prime 0}(\bar{q^\prime}_{L} \Phi q_{R}^\prime + c.c.) 
-m_{\Phi 0}^2\tr(\Phi^\dagger\Phi) 
\end{align}
is renormalized as 
\begin{eqnarray}
\mathcal{L}_0 &=&
\bar{q^\prime}i\Slash{\partial} q^\prime -y_{q^\prime 0}(\bar{q^\prime}_{L} \Phi q_{R}^\prime + c.c.) 
\nonumber\\
&+&
Z_\Phi \tr (\partial _\mu \Phi ^\dagger \partial ^\mu \Phi ) - m_{\Phi}^{2} \tr\Phi ^\dagger \Phi \\
&-& 
  \frac{\lambda _1}{2} (\tr\Phi ^\dagger \Phi)^2 - \frac{\lambda _2}{2}\tr (\Phi ^\dagger \Phi )^2 
+ c(\det\Phi+c.c.)  \nonumber
\end{eqnarray}
in terms of the bare field variables, where
\begin{eqnarray}
Z_\Phi&=&\frac{N_c}{16\pi^2}y_{q^\prime 0}^2\ln(\Lambda_{\rm 4f}^2/\mu^2),  \\
m_\Phi^2&=&m_{\Phi 0}^2-\frac{2N_c}{16\pi^2}y_{q^\prime 0}^2(\Lambda_{\rm 4f}^2-\mu^2), \\
\lambda_2 &=&\frac{2N_c}{16\pi^2}y_{q^\prime 0}^4\ln(\Lambda_{\rm 4f}^2/\mu^2), \\
\lambda_1&=&0 ,  \qquad c=0 . 
\end{eqnarray}
Then, the renormalized couplings read
\begin{eqnarray}
\bar y(\mu)^2 &=&  \frac{16\pi^2}{N_c} \frac{1}{\ln(\Lambda_{\rm 4f}^2/\mu^2)}, 
\nonumber\\
\bar \lambda_1(\mu) &=&  0,  \\
\bar \lambda_2(\mu) &=& \frac{32\pi^2}{N_c} \frac{1}{\ln(\Lambda_{\rm 4f}^2/\mu^2)} .  
\end{eqnarray}


\begin{thebibliography}{99}

\bibitem{Sakharov:1973}
A.D.~Sakharov, Pis'ma Zh. Eksp. Teor. Fiz. {\bf 5}, 32 (1967)  [JETP Lett. {\bf 5}, 24 (1967)]

\bibitem{Kuzmin:1985mm}
  V.~A.~Kuzmin, V.~A.~Rubakov and M.~E.~Shaposhnikov,
  Phys.\ Lett.\  B {\bf 155}, 36 (1985).
  
  
\bibitem{Cohen:1990py}
  A.~G.~Cohen, D.~B.~Kaplan and A.~E.~Nelson,
  Phys.\ Lett.\  B {\bf 245}, 561 (1990).
  
\bibitem{Cohen:1990it}
  A.~G.~Cohen, D.~B.~Kaplan and A.~E.~Nelson,
  Nucl.\ Phys.\  B {\bf 349}, 727 (1991).

  
\bibitem{Steigman:2005uz}
  G.~Steigman,
  Int.\ J.\ Mod.\ Phys.\  E {\bf 15}, 1 (2006)
  [arXiv:astro-ph/0511534].
  
  
\bibitem{Spergel:2006hy}
  D.~N.~Spergel {\it et al.}  [WMAP Collaboration],
  arXiv:astro-ph/0603449.

  
 
\bibitem{Kobayashi:1973fv}
  M.~Kobayashi and T.~Maskawa,
  Prog.\ Theor.\ Phys.\  {\bf 49}, 652 (1973).

%


\bibitem{Shaposhnikov:1987tw}
  M.~E.~Shaposhnikov,
  Nucl.\ Phys.\ B {\bf 287}, 757 (1987).

\bibitem{Farrar:1993sp}
  G.~R.~Farrar and M.~E.~Shaposhnikov,
  Phys.\ Rev.\ Lett.\  {\bf 70}, 2833 (1993)
  [Erratum-ibid.\  {\bf 71}, 210 (1993)]
  [arXiv:hep-ph/9305274].

\bibitem{Farrar:1993hn}
  G.~R.~Farrar and M.~E.~Shaposhnikov,
  Phys.\ Rev.\ D {\bf 50}, 774 (1994)
  [arXiv:hep-ph/9305275].

\bibitem{Gavela:1994dt}
  M.~B.~Gavela, P.~Hernandez, J.~Orloff, O.~Pene and C.~Quimbay,
  Nucl.\ Phys.\ B {\bf 430}, 382 (1994)
  [arXiv:hep-ph/9406289].

\bibitem{Gavela:1993ts}
  M.~B.~Gavela, P.~Hernandez, J.~Orloff and O.~Pene,
  Mod.\ Phys.\ Lett.\ A {\bf 9}, 795 (1994)
  [arXiv:hep-ph/9312215].

\bibitem{Huet:1994jb}
  P.~Huet and E.~Sather,
  Phys.\ Rev.\ D {\bf 51}, 379 (1995)
  [arXiv:hep-ph/9404302].


\bibitem{:2003ih}
    [LEP Collaboration],
  arXiv:hep-ex/0312023.

\bibitem{Kajantie:1996mn}
  K.~Kajantie, M.~Laine, K.~Rummukainen and M.~E.~Shaposhnikov,
  Phys.\ Rev.\ Lett.\  {\bf 77}, 2887 (1996)
  [arXiv:hep-ph/9605288].

\bibitem{Rummukainen:1998as}
  K.~Rummukainen, M.~Tsypin, K.~Kajantie, M.~Laine and M.~E.~Shaposhnikov,
  Nucl.\ Phys.\ B {\bf 532}, 283 (1998)
  [arXiv:hep-lat/9805013].

\bibitem{Csikor:1998eu}
  F.~Csikor, Z.~Fodor and J.~Heitger,
  Phys.\ Rev.\ Lett.\  {\bf 82}, 21 (1999)
  [arXiv:hep-ph/9809291].
  
\bibitem{Aoki:1999fi}
  Y.~Aoki, F.~Csikor, Z.~Fodor and A.~Ukawa,
  Phys.\ Rev.\  D {\bf 60}, 013001 (1999)
  [arXiv:hep-lat/9901021].
  



\bibitem{Alcaraz:2006mx}
  J.~Alcaraz {\it et al.}  [ALEPH Collaboration and DELPHI Collaboration and
                  L3 Collaboration and ],
  arXiv:hep-ex/0612034.
  
\bibitem{Holdom:1990tc}
  B.~Holdom and J.~Terning,
  Phys.\ Lett.\  B {\bf 247}, 88 (1990).

\bibitem{Peskin:1990zt}
  M.~E.~Peskin and T.~Takeuchi,
  Phys.\ Rev.\ Lett.\  {\bf 65}, 964 (1990).
  
\bibitem{Peskin:1991sw}
  M.~E.~Peskin and T.~Takeuchi,
  Phys.\ Rev.\  D {\bf 46}, 381 (1992).

\bibitem{Golden:1990ig}
  M.~Golden and L.~Randall,
  Nucl.\ Phys.\  B {\bf 361}, 3 (1991).

 
\bibitem{Gates:1991uu}
  E.~Gates and J.~Terning,
  Phys.\ Rev.\ Lett.\  {\bf 67}, 1840 (1991).

\bibitem{Kniehl:1992ez}
  B.~A.~Kniehl and H.~G.~Kohrs,
  Phys.\ Rev.\  D {\bf 48}, 225 (1993).
  
\bibitem{Holdom:1996bn}
  B.~Holdom,
  Phys.\ Rev.\  D {\bf 54}, 721 (1996)
  [arXiv:hep-ph/9602248].
 
\bibitem{He:2001tp}
  H.~J.~He, N.~Polonsky and S.~f.~Su,
  Phys.\ Rev.\  D {\bf 64}, 053004 (2001)
  [arXiv:hep-ph/0102144].
  
  
\bibitem{Kribs:2007nz}
  G.~D.~Kribs, T.~Plehn, M.~Spannowsky and T.~M.~P.~Tait,
  Phys.\ Rev.\  D {\bf 76}, 075016 (2007)
  [arXiv:0706.3718 [hep-ph]].


\bibitem{Hou:2005hd}
  W.~S.~Hou, M.~Nagashima and A.~Soddu,
  Phys.\ Rev.\ Lett.\  {\bf 95}, 141601 (2005)
  [arXiv:hep-ph/0503072].
  
  
\bibitem{Hou:2006zza}
  W.~S.~Hou, M.~Nagashima, G.~Raz and A.~Soddu,
  JHEP {\bf 0609}, 012 (2006)
  [arXiv:hep-ph/0603097].

\bibitem{Hou:2006jy}
  W.~S.~Hou, H.~n.~Li, S.~Mishima and M.~Nagashima,
  Phys.\ Rev.\ Lett.\  {\bf 98}, 131801 (2007)
  [arXiv:hep-ph/0611107].

\bibitem{Soni:2008bc}
  A.~Soni, A.~K.~Alok, A.~Giri, R.~Mohanta and S.~Nandi,
  arXiv:0807.1971 [hep-ph].
  
\bibitem{Hou:2008xd}
  W.~S.~Hou,
  arXiv:0803.1234 [hep-ph].

\bibitem{Fok:2008yg}
  R.~Fok and G.~D.~Kribs,
  arXiv:0803.4207 [hep-ph].

\bibitem{Carena:2004ha}
  M.~Carena, A.~Megevand, M.~Quiros and C.~E.~M.~Wagner,
  Nucl.\ Phys.\  B {\bf 716}, 319 (2005)
  [arXiv:hep-ph/0410352].

\bibitem{Ham:2004xh}
  S.~W.~Ham, S.~K.~Oh and D.~Son,
  Phys.\ Rev.\  D {\bf 71}, 015001 (2005)
  [arXiv:hep-ph/0411012].

\bibitem{Chanowitz:1978mv}
  M.~S.~Chanowitz, M.~A.~Furman and I.~Hinchliffe,
  Nucl.\ Phys.\  B {\bf 153}, 402 (1979).
  
\bibitem{Nambu:1961tp}
  Y.~Nambu and G.~Jona-Lasinio,
  Phys.\ Rev.\  {\bf 122} (1961) 345.

  
\bibitem{Nambu:1988mr}
  Y.~Nambu, ``Quasisupersymmetry, bootstrap symmetry breaking and fermion masses", Invited talk, in: M. Bando, T. Muta, Y. Yamawaki (Eds.), Proceedings of 1988 International Workshop New Trends in Strong Coupling Gauge Theories, Nagoya, Japan, August 24--27, 1988, World Scientific, Singapore, 1989.
  
    
\bibitem{Miransky:1988xi}
  V.~A.~Miransky, M.~Tanabashi and K.~Yamawaki,
  Phys.\ Lett.\  B {\bf 221}, 177 (1989).

\bibitem{Miransky:1989ds}
  V.~A.~Miransky, M.~Tanabashi and K.~Yamawaki,
  Mod.\ Phys.\ Lett.\  A {\bf 4}, 1043 (1989).
  
\bibitem{Marciano:1989mj}
  W.~J.~Marciano,
  Phys.\ Rev.\  D {\bf 41}, 219 (1990).
  
\bibitem{Marciano:1989xd}
  W.~J.~Marciano,
  Phys.\ Rev.\ Lett.\  {\bf 62}, 2793 (1989).
      
\bibitem{Bardeen:1989ds}
  W.~A.~Bardeen, C.~T.~Hill and M.~Lindner,
  Phys.\ Rev.\  D {\bf 41}, 1647 (1990).
  

\bibitem{Holdom:1986rn}
  B.~Holdom,
  Phys.\ Rev.\ Lett.\  {\bf 57}, 2496 (1986)
  [Erratum-ibid.\  {\bf 58}, 177 (1987)].

\bibitem{Holdom:1983kw}
  B.~Holdom,
  Phys.\ Lett.\  B {\bf 143}, 227 (1984).

\bibitem{Holdom:1990ta}
  B.~Holdom,
  Phys.\ Lett.\  B {\bf 246}, 169 (1990).
  
\bibitem{Hill:1990ge}
  C.~T.~Hill, M.~A.~Luty and E.~A.~Paschos,
  Phys.\ Rev.\  D {\bf 43}, 3011 (1991).
  
\bibitem{Elliott:1992xg}
  T.~Elliott and S.~F.~King,
  Phys.\ Lett.\  B {\bf 283}, 371 (1992).
  
\bibitem{Hill:1992ev}
  C.~T.~Hill, D.~C.~Kennedy, T.~Onogi and H.~L.~Yu,
  Phys.\ Rev.\  D {\bf 47}, 2940 (1993)
  [arXiv:hep-ph/9210233].
  
  
\bibitem{Coleman:1973jx}
  S.~R.~Coleman and E.~Weinberg,
  Phys.\ Rev.\  D {\bf 7}, 1888 (1973).


\bibitem{Bak-Krinsky-Mukamel:1976}
P.~Bak, S.~Krinsky and D.~Mukamel, Phys.\ Rev.\ Lett.\  {\bf 36}, 52 (1976)
 
\bibitem{Rudnick:1978}
J.~Rudnic, Phys.\ Rev.\ B {\bf 18}, 1406 (1978). 

\bibitem{Amit:1978dk}
  D.~J.~Amit,
``Field Theory, The Renormalization Group, And Critical Phenomena,''
revised second edition, {\it  World Scientific  1984, 394p}

\bibitem{Ginsparg:1980ef}
  P.~H.~Ginsparg,
  Nucl.\ Phys.\  B {\bf 170}, 388 (1980).

\bibitem{Iacobson:1981jm}
  H.~H.~Iacobson and D.~J.~Amit,
  Annals Phys.\  {\bf 133}, 57 (1981).

\bibitem{Yamagishi:1981qq}
  H.~Yamagishi,
  Phys.\ Rev.\  D {\bf 23} (1981) 1880.

\bibitem{Pisarski:1983ms}
  R.~D.~Pisarski and F.~Wilczek,
  Phys.\ Rev.\ D {\bf 29}, 338 (1984).

  
\bibitem{Chivukula:1992pm}
  R.~S.~Chivukula, M.~Golden and E.~H.~Simmons,
  Phys.\ Rev.\ Lett.\  {\bf 70}, 1587 (1993)
  [arXiv:hep-ph/9210276].
  


\bibitem{Grojean:2004xa}
  C.~Grojean, G.~Servant and J.~D.~Wells,
  Phys.\ Rev.\ D {\bf 71}, 036001 (2005)
  [arXiv:hep-ph/0407019].
  
\bibitem{Ham:2004zs}
  S.~W.~Ham and S.~K.~Oh,
  Phys.\ Rev.\ D {\bf 70}, 093007 (2004)
  [arXiv:hep-ph/0408324].
  
\bibitem{Bodeker:2004ws}
  D.~Bodeker, L.~Fromme, S.~J.~Huber and M.~Seniuch,
  JHEP {\bf 0502}, 026 (2005)
  [arXiv:hep-ph/0412366].

\bibitem{Delaunay:2007wb}
  C.~Delaunay, C.~Grojean and J.~D.~Wells,
  JHEP {\bf 0804}, 029 (2008)
  [arXiv:0711.2511 [hep-ph]].

\bibitem{Grinstein:2008qi}
  B.~Grinstein and M.~Trott,
  Phys.\ Rev.\  D {\bf 78}, 075022 (2008)
  [arXiv:0806.1971 [hep-ph]].


 
\bibitem{Kikukawa:2007zk}
  Y.~Kikukawa, M.~Kohda and J.~Yasuda,
  Phys.\ Rev.\  D {\bf 77}, 015014 (2008)
  [arXiv:0709.2221 [hep-ph]].
  
\bibitem{Cline:2008hr}
  J.~M.~Cline, M.~Jarvinen and F.~Sannino,
  Phys.\ Rev.\  D {\bf 78}, 075027 (2008)
  [arXiv:0808.1512 [hep-ph]].

\bibitem{Jarvinen:2009wr}
  M.~Jarvinen, T.~A.~Ryttov and F.~Sannino,
  arXiv:0901.0496 [hep-ph].
  
\bibitem{Espinosa:2004pn}
  J.~R.~Espinosa, M.~Losada and A.~Riotto,
  Phys.\ Rev.\  D {\bf 72}, 043520 (2005)
  [arXiv:hep-ph/0409070].
  



\bibitem{Panico:2005ft}
  G.~Panico and M.~Serone,
  JHEP {\bf 0505}, 024 (2005)
  [arXiv:hep-ph/0502255].

\bibitem{Maru:2005jy}
  N.~Maru and K.~Takenaga,
  Phys.\ Rev.\  D {\bf 72}, 046003 (2005)
  [arXiv:hep-th/0505066].

\bibitem{Maru:2006wx}
  N.~Maru and K.~Takenaga,
  Phys.\ Rev.\  D {\bf 74}, 015017 (2006)
  [arXiv:hep-ph/0606139].
  
  

\bibitem{Holdom:2006mr}
  B.~Holdom,
  JHEP {\bf 0608}, 076 (2006)
  [arXiv:hep-ph/0606146].
  



\bibitem{Dolan:1973qd}
  L.~Dolan and R.~Jackiw,
  Phys.\ Rev.\  D {\bf 9}, 3320 (1974).
  
\bibitem{Weinberg:1974hy}
  S.~Weinberg,
  Phys.\ Rev.\  D {\bf 9}, 3357 (1974).

\bibitem{Anderson:1991zb}
  G.~W.~Anderson and L.~J.~Hall,
  Phys.\ Rev.\  D {\bf 45}, 2685 (1992).
 

\bibitem{Fendley:1987ef}
  P.~Fendley,
  Phys.\ Lett.\  B {\bf 196}, 175 (1987).

\bibitem{Espinosa:1992gq}
  J.~R.~Espinosa, M.~Quiros and F.~Zwirner,
  Phys.\ Lett.\  B {\bf 291}, 115 (1992)
  [arXiv:hep-ph/9206227].

\bibitem{Parwani:1991gq}
  R.~R.~Parwani,
  Phys.\ Rev.\  D {\bf 45}, 4695 (1992)
  [Erratum-ibid.\  D {\bf 48}, 5965 (1993)]
  [arXiv:hep-ph/9204216].

 
\bibitem{Carrington:1991hz}
  M.~E.~Carrington,
  Phys.\ Rev.\  D {\bf 45}, 2933 (1992).

\bibitem{Dine:1992wr}
  M.~Dine, R.~G.~Leigh, P.~Y.~Huet, A.~D.~Linde and D.~A.~Linde,
  Phys.\ Rev.\  D {\bf 46}, 550 (1992)
  [arXiv:hep-ph/9203203].
  
\bibitem{Arnold:1992rz}
  P.~Arnold and O.~Espinosa,
  Phys.\ Rev.\  D {\bf 47}, 3546 (1993)
  [Erratum-ibid.\  D {\bf 50}, 6662 (1994)]
  [arXiv:hep-ph/9212235].

\bibitem{Fodor:1994bs}
  Z.~Fodor and A.~Hebecker,
  Nucl.\ Phys.\  B {\bf 432}, 127 (1994)
  [arXiv:hep-ph/9403219].


\bibitem{Harada:1994wy}
  M.~Harada, Y.~Kikukawa, T.~Kugo and H.~Nakano,
  Prog.\ Theor.\ Phys.\  {\bf 92}, 1161 (1994)
  [arXiv:hep-ph/9407398].
  
  

\bibitem{Bochkarev:1990fx}
  A.~I.~Bochkarev, S.~V.~Kuzmin and M.~E.~Shaposhnikov,
  Phys.\ Lett.\  B {\bf 244}, 275 (1990).

\bibitem{Bochkarev:1990gb}
  A.~I.~Bochkarev, S.~V.~Kuzmin and M.~E.~Shaposhnikov,
  Phys.\ Rev.\  D {\bf 43}, 369 (1991).

\bibitem{Cohen:1991iu}
  A.~G.~Cohen, D.~B.~Kaplan and A.~E.~Nelson,
  Phys.\ Lett.\  B {\bf 263}, 86 (1991).

\bibitem{Nelson:1991ab}
  A.~E.~Nelson, D.~B.~Kaplan and A.~G.~Cohen,
  Nucl.\ Phys.\  B {\bf 373}, 453 (1992).


\bibitem{Turok:1990zg}
  N.~Turok and J.~Zadrozny,
  Nucl.\ Phys.\  B {\bf 358}, 471 (1991).
%

\bibitem{Turok:1991uc}
  N.~Turok and J.~Zadrozny,
  Nucl.\ Phys.\  B {\bf 369}, 729 (1992).
  
  
\bibitem{Funakubo:1993jg}
  K.~Funakubo, A.~Kakuto and K.~Takenaga,
  Prog.\ Theor.\ Phys.\  {\bf 91}, 341 (1994)
  [arXiv:hep-ph/9310267].

\bibitem{Davies:1994id}
  A.~T.~Davies, C.~D.~Froggatt, G.~Jenkins and R.~G.~Moorhouse,
  Phys.\ Lett.\  B {\bf 336}, 464 (1994).

\bibitem{Cline:1995dg}
  J.~M.~Cline, K.~Kainulainen and A.~P.~Vischer,
  Phys.\ Rev.\  D {\bf 54}, 2451 (1996)
  [arXiv:hep-ph/9506284].
  
\bibitem{Cline:1996mg}
  J.~M.~Cline and P.~A.~Lemieux,
  Phys.\ Rev.\ D {\bf 55}, 3873 (1997)
  [arXiv:hep-ph/9609240].
  
\bibitem{Fromme:2006cm}
  L.~Fromme, S.~J.~Huber and M.~Seniuch,
  JHEP {\bf 0611}, 038 (2006)
  [arXiv:hep-ph/0605242].
  

\bibitem{Kanemura:2004ch}
  S.~Kanemura, Y.~Okada and E.~Senaha,
  Phys.\ Lett.\  B {\bf 606}, 361 (2005)
  [arXiv:hep-ph/0411354].

\bibitem{Aoki:2008av}
  M.~Aoki, S.~Kanemura and O.~Seto,
  arXiv:0807.0361 [hep-ph].

   
  
\bibitem{Kikukawa-Kohda:2009xx}
Y.~Kikukawa, M.~Kohda and J.~Yasuda, in preparation.


 
  
\bibitem{Hill:1980sq}
  C.~T.~Hill,
  Phys.\ Rev.\  D {\bf 24}, 691 (1981).

\bibitem{Hill:1985tg}
  C.~T.~Hill, C.~N.~Leung and S.~Rao,
  Nucl.\ Phys.\  B {\bf 262}, 517 (1985).
  


\bibitem{Manton:1983nd}
  N.~S.~Manton,
  Phys.\ Rev.\  D {\bf 28}, 2019 (1983).
  
\bibitem{Klinkhamer:1984di}
  F.~R.~Klinkhamer and N.~S.~Manton,
  Phys.\ Rev.\  D {\bf 30}, 2212 (1984).
  
\bibitem{Arnold:1987mh}
  P.~Arnold and L.~D.~McLerran,
  Phys.\ Rev.\  D {\bf 36}, 581 (1987).

\bibitem{Akiba:1989xu}
  T.~Akiba, H.~Kikuchi and T.~Yanagida,
  Phys.\ Rev.\  D {\bf 40}, 588 (1989).
  
  
\bibitem{Carson:1990jm}
  L.~Carson, X.~Li, L.~D.~McLerran and R.~T.~Wang,
  Phys.\ Rev.\  D {\bf 42}, 2127 (1990).
  
  
  

\bibitem{Baacke:1993aj}
  J.~Baacke and S.~Junker,
  Phys.\ Rev.\  D {\bf 49}, 2055 (1994)
  [arXiv:hep-ph/9308310].

\bibitem{Baacke:1994ix}
  J.~Baacke and S.~Junker,
  Phys.\ Rev.\  D {\bf 50}, 4227 (1994)
  [arXiv:hep-th/9402078].
  
\bibitem{Moore:1995jv}
  G.~D.~Moore,
  Phys.\ Rev.\  D {\bf 53}, 5906 (1996)
  [arXiv:hep-ph/9508405].
   
\bibitem{Kovner:1999ja}
  A.~Kovner, A.~Krasnitz and R.~Potting,
  Phys.\ Rev.\  D {\bf 61}, 025009 (2000)
  [arXiv:hep-ph/9907381].
  

\bibitem{Amsler:2008zz}
  C.~Amsler {\it et al.}  [Particle Data Group],
  Phys.\ Lett.\  B {\bf 667}, 1 (2008).
  
\bibitem{Hasenfratz:1991it}
  A.~Hasenfratz, P.~Hasenfratz, K.~Jansen, J.~Kuti and Y.~Shen,
  Nucl.\ Phys.\  B {\bf 365}, 79 (1991).
  

\bibitem{Neuberger:1997fp}
  H.~Neuberger,
  Phys.\ Lett.\  B {\bf 417}, 141 (1998)
  [arXiv:hep-lat/9707022].
  
\bibitem{Neuberger:1998wv}
  H.~Neuberger,
  Phys.\ Lett.\  B {\bf 427}, 353 (1998)
  [arXiv:hep-lat/9801031].

\bibitem{Kikukawa:1997qh}
  Y.~Kikukawa and H.~Neuberger,
  Nucl.\ Phys.\  B {\bf 513}, 735 (1998)
  [arXiv:hep-lat/9707016].
     
\bibitem{Ginsparg:1981bj}
  P.~H.~Ginsparg and K.~G.~Wilson,
  Phys.\ Rev.\  D {\bf 25}, 2649 (1982).

\bibitem{Luscher:1998pqa}
  M.~Luscher,
  Phys.\ Lett.\  B {\bf 428}, 342 (1998)
  [arXiv:hep-lat/9802011].
  
\bibitem{Luscher:1998du}
  M.~Luscher,
  Nucl.\ Phys.\  B {\bf 549}, 295 (1999)
  [arXiv:hep-lat/9811032].
  
\bibitem{Kadoh:2007xb}
  D.~Kadoh and Y.~Kikukawa,
  JHEP {\bf 0805}, 095 (2008)
  [arXiv:0709.3658 [hep-lat]].
  

\bibitem{Neuberger:1998my}
  H.~Neuberger,
  Phys.\ Rev.\ Lett.\  {\bf 81}, 4060 (1998)
  [arXiv:hep-lat/9806025].
  
\bibitem{Frommer:1995ik}
  A.~Frommer, B.~Nockel, S.~Gusken, T.~Lippert and K.~Schilling,
  Int.\ J.\ Mod.\ Phys.\  C {\bf 6}, 627 (1995)
  [arXiv:hep-lat/9504020].

\bibitem{van den Eshof:2002ms}
  J.~van den Eshof, A.~Frommer, T.~Lippert, K.~Schilling and H.~A.~van der Vorst,
  Comput.\ Phys.\ Commun.\  {\bf 146}, 203 (2002)
  [arXiv:hep-lat/0202025].

\bibitem{Cundy:2004pza}
  N.~Cundy, J.~van den Eshof, A.~Frommer, S.~Krieg, T.~Lippert and K.~Schafer,
  Comput.\ Phys.\ Commun.\  {\bf 165}, 221 (2005)
  [arXiv:hep-lat/0405003].

\bibitem{Chiu:2002eh}
  T.~W.~Chiu, T.~H.~Hsieh, C.~H.~Huang and T.~R.~Huang,
  Phys.\ Rev.\  D {\bf 66}, 114502 (2002)
  [arXiv:hep-lat/0206007].
  

\bibitem{Vranas:2006zk}
  P.~M.~Vranas,
  Phys.\ Rev.\  D {\bf 74}, 034512 (2006)
  [arXiv:hep-lat/0606014].
  
\bibitem{Fukaya:2006vs}
  H.~Fukaya, S.~Hashimoto, K.~I.~Ishikawa, T.~Kaneko, H.~Matsufuru, T.~Onogi and N.~Yamada
                  [JLQCD Collaboration],
  Phys.\ Rev.\  D {\bf 74}, 094505 (2006)
  [arXiv:hep-lat/0607020].

\cite{Fukaya:2007fb}
\bibitem{Fukaya:2007fb}
  H.~Fukaya {\it et al.}  [JLQCD Collaboration],
  Phys.\ Rev.\ Lett.\  {\bf 98}, 172001 (2007)
  [arXiv:hep-lat/0702003].

\bibitem{Aoki:2008tq}
  S.~Aoki {\it et al.}  [JLQCD Collaboration],
  Phys.\ Rev.\  D {\bf 78}, 014508 (2008)
  [arXiv:0803.3197 [hep-lat]].

 
\bibitem{Gerhold:2007yb}
  P.~Gerhold and K.~Jansen,
  JHEP {\bf 0709}, 041 (2007)
  [arXiv:0705.2539 [hep-lat]].

\bibitem{Gerhold:2007gx}
  P.~Gerhold and K.~Jansen,
  JHEP {\bf 0710}, 001 (2007)
  [arXiv:0707.3849 [hep-lat]].
 
  
 
\bibitem{Gerhold:2008mb}
  P.~Gerhold, K.~Jansen and J.~Kallarackal,
  arXiv:0810.4447 [hep-lat].
  
\end{thebibliography}

\end{document}